\documentclass[fleqn,usenatbib]{mnras}
\usepackage{newtxtext,newtxmath}

\usepackage[T1]{fontenc}
\usepackage{ae,aecompl}
\usepackage{graphicx}	
\usepackage{amsmath}	
\usepackage{enumitem}   
\usepackage{natbib}

\defcitealias{2020dudz}{D20}

\title[Tracing the dust evolution using STUDIES \& AS2UDS]{Tracing the evolution of dust-obscured activity using sub-millimetre galaxy populations from STUDIES and AS2UDS}

\author[U.\ Dudzevi\v{c}i\={u}t\.{e} et al.]{U.\ Dudzevi\v{c}i\={u}t\.{e},$^{1}$\thanks{E-mail: ugne.dudzeviciute2@durham.ac.uk}
Ian Smail,$^{1}$
A.\,M.\ Swinbank,$^{1}$
C.-F.\ Lim,$^{2,3}$
W.-H.\ Wang,$^{3}$
\newauthor J.\,M.\ Simpson,$^{1}$
Y.\ Ao,$^{4}$
S.\,C.\ Chapman,$^{5,6,7}$
C.-C.\ Chen,$^{3,8}$
D.\ Clements,$^{9}$
\newauthor H.\ Dannerbauer,$^{10,11}$
L.\,C.\ Ho,$^{12,13}$
H.\,S.\ Hwang,$^{14}$
M.\ Koprowski,$^{15}$
C.-H.\ Lee,$^{16}$
\newauthor D.\ Scott,$^{5}$
H.\ Shim,$^{17}$
R.\ Shirley,$^{10,11}$ and
Y.\ Toba$^{3,18,19}$\\
$^{1}$ Centre for Extragalactic Astronomy, Department of Physics, Durham University, South Road, Durham DH1 3LE, UK\\
$^{2}$ Graduate Institute of Astrophysics, National Taiwan University, Taipei 10617, Taiwan\\
$^{3}$ Academia Sinica Institute of Astronomy and Astrophysics (ASIAA), No.\ 1, Sec.\ 4, Roosevelt Rd., Taipei 10617, Taiwan\\
$^{4}$ Purple Mountain Observatory, Chinese Academy of Sciences, Nanjing 210033, People's Republic of China\\
$^{5}$ Department of Physics and Astronomy, University of British Columbia, 6225 Agricultural Road, Vancouver, BC V6T 1Z1, Canada\\
$^{6}$ National Research Council, Herzberg Astronomy and Astrophysics, 5071 West Saanich Road, Victoria, BC V9E 2E7, Canada\\
$^{7}$ Department of Physics and Atmospheric Science, Dalhousie University, Halifax, NS B3H 4R2, Canada\\
$^{8}$ European Southern Observatory, Karl Schwarzschild Strasse 2, Garching, Germany\\
$^{9}$ Blackett Lab, Imperial College, London, Prince Consort Road, London SW7 2AZ, UK\\
$^{10}$ Instituto de Astrof\'isica de Canarias, E-38205 La Laguna, Tenerife, Spain\\
$^{11}$ Universidad de La Laguna, Departamento de Astrof\'isica, E-38206 La Laguna, Tenerife, Spain\\
$^{12}$ Kavli Institute for Astronomy and Astrophysics, Peking University, Beijing 100871, People's Republic of China\\
$^{13}$ Department of Astronomy, School of Physics, Peking University, Beijing 100871, People's Republic of China\\
$^{14}$ Korea Astronomy and Space Science Institute, 776 Daedeokdae-ro, Yuseong-gu, Daejeon 34055, Republic of Korea\\
$^{15}$ Institute of Astronomy, Faculty of Physics, Astronomy and Informatics, Nicolaus Copernicus University, Grudziadzka 5,\\ 87-100 Torun, Poland\\
$^{16}$ NSF's National Optical-Infrared Astronomy Research Laboratory, 950 North Cherry Avenue, Tucson, AZ 85719, USA\\
$^{17}$ Department of Earth Science Education, Kyungpook National University, Daegu 41566, Republic of Korea\\
$^{18}$ Department of Astronomy, Kyoto University, Kitashirakawa-Oiwake-cho, Sakyo-ku, Kyoto 606-8502, Japan\\
$^{19}$ Research Center for Space and Cosmic Evolution, Ehime University, 2-5 Bunkyo-cho, Matsuyama, Ehime 790-8577, Japan
}

\date{Accepted XXX. Received YYY; in original form ZZZ}

\pubyear{2020}
\begin{document}
\label{firstpage}
\pagerange{\pageref{firstpage}--\pageref{lastpage}}
\maketitle
\begin{abstract}
We analyse the physical properties of 121 SNR\,$\geq$\,5 sub-millimetre galaxies (SMGs) from the STUDIES 450-$\mu$m survey. We model their UV-to-radio spectral energy distributions using {\sc magphys}+photo-$z$ and compare the results to similar modelling of 850-$\mu$m-selected SMG sample from AS2UDS, to understand the fundamental physical differences between the two populations at the observed depths.
The redshift distribution of the 450-$\mu$m sample has a median of $z$\,$=$\,1.85\,$\pm$\,0.12 and can be described by strong evolution of the far-infrared luminosity function.
The fainter 450-$\mu$m sample has $\sim$\,14 times higher space density than the brighter 850-$\mu$m sample at $z$\,$\lesssim$\,2, and a comparable space density at $z$\,$=$\,2--3, before rapidly declining, suggesting LIRGs are the main obscured population at $z$\,$\sim$\,1--2, while ULIRGs dominate at higher redshifts. 
We construct rest-frame $\sim$\,180-$\mu$m-selected and dust-mass-matched samples at $z$\,$=$\,1--2 and $z$\,$=$\,3--4 from the 450-$\mu$m and 850-$\mu$m samples, respectively, to probe the evolution of a uniform sample of galaxies spanning the cosmic noon era.
Using far-infrared luminosity, dust masses and an optically-thick dust model, we suggest that higher-redshift sources have higher dust densities due to inferred dust continuum sizes which are roughly half of those for the
lower-redshift population at a given dust mass, leading to higher dust attenuation. 
We track the evolution in the cosmic dust mass density and suggest that the dust content of galaxies is governed by a combination of both the variation of gas content and dust destruction timescale.
\end{abstract}


\section{Introduction} \label{intro}

The discovery of a significant  energy density in the extragalactic background of the Universe at wavelengths $\geq$\,150\,$\mu$m \citep{1996puget,1998fixsen}, suggested the existence of a population of dust-enshrouded galaxies that is much more significant than their local analogues (\citealt{1968low&kleinmann,1972rieke&low,1984neugebauer}; see \citealt{2001hauser&dwek} and \citealt{2005lagache} for reviews). In such far-infrared luminous galaxies, the radiation from young massive stars is absorbed by  dust grains in their interstellar medium (ISM) and re-radiated as thermal continuum emission at far-infrared wavelengths \citep[for reviews, see][]{2014lutz,2014casey}. In the mid-1990s, the first surveys at sub-millimetre  wavelengths (450\,$\mu$m and 850\,$\mu$m) using the Sub-millimeter Common User Bolometric Array (SCUBA) on James Clerk Maxwell Telescope (JCMT) began to resolve this far-infrared/sub-millimetre background into its constituent galaxies and identified the first statistical samples of  high-redshift, sub-millimetre bright galaxies (SMGs -- \citealt{1997smail,1998hughes,1998barger,1999eales}). These surveys  confirmed the cosmological significance of far-infrared-luminous galaxies, in particular their potentially significant contribution to the star-formation rate density at high redshifts \citep[see][]{2014madau}.

Due to atmospheric transmission, large-scale surveys of the  high-redshift SMG populations are undertaken primarily in wavebands around 850\,$\mu$m and 1.2\,mm \citep[e.g.][]{2006coppin,2008scott,2009weiss,2011hatsukade,2013mocanu,2014umehata,2017geach,2017miettinen,2018cowie,2019stach}. 
These wavebands typically select galaxies based on their luminosity at rest-frame wavelengths  around 300\,$\mu$m and are thus sensitive to the cool dust mass of the galaxies (\citealt{2020dudz}, hereafter \citetalias{2020dudz}).
Subsequent studies of this population have suggested that these systems are strongly dust obscured systems with high far-infrared luminosities and lying at high redshifts, with a number density peaking at $z$\,$\sim$\,2--3 (\citealt{2005chapman,2015dacunha,2017danielson,2016koprowski,2017brisbin}; \citetalias{2020dudz}). SMGs have huge gas reservoirs of the order of 10$^{10-11}$\,M$_\odot$ and star-formation rates ranging over 100--1000\,M$_\odot$\,yr$^{-1}$, meaning that they have the capacity to rapidly increase their already high stellar masses ($\sim$\,10$^{11}$\,M$_\odot$) on a short timescale (\citealt{2011ivison,2013bothwell}; \citetalias{2020dudz}). Although these  studies have provided insight into the physical properties of the SMGs, observations at such long wavelengths do not sample the peak of the far-infrared emission and, as noted, are more sensitive to sources with larger masses of cool dust, as well as those at higher redshifts \citep{1993blain&longair}. 

 At high redshift ($z$\,$>$\,2), selection at shorter submillimetre and far-infrared wavelengths (e.g.\ $\leq$\,500\,$\mu$m) samples the spectral energy distribution (SED) of the dust continuum emission closer to the peak of the far-infrared emission, rather than the Rayleigh-Jeans tail which is traced by the 850\,$\mu$m and 1.2\,mm surveys. Therefore, surveys at
 shorter wavelengths are more sensitive to  far-infrared luminosity, than the cold dust mass. This can potentially lead to surveys detecting physically different sources when selected at different wavebands and redshifts.
 
Surveys at far-infrared wavelengths, specifically 250, 350 and 500\,$\mu$m, using the SPIRE instrument on {\it Herschel} have mapped hundreds of square degrees of sky \citep{2010eales,2012oliver,2014wang,2016valiante}.
While covering huge areas, these surveys are limited in sensitivity due to the large beam size and resulting bright confusion limit\footnote{Defined as the sensitivity limit arising from unresolved sources which cannot be improved by increasing the integration time.}, which makes it challenging to detect all but the brightest (unlensed) sources at $z$\,$\gtrsim$\,1 (\citealt{2011symeonidis}, although see \citealt{2016shu,2018jin,2018liu}). Moreover, the large beam size makes it difficult to reliably locate counterparts needed to understand their properties. However, higher-resolution imaging can be obtained from single-dish telescopes on the ground through the atmospheric windows at 350\,$\mu$m \citep{2007khan,2008coppin} and 450\,$\mu$m \citep{1999blain,2013chen}. Unfortunately the atmospheric transmission at 450\,$\mu$m is around half of that at 850\,$\mu$m, and obtaining deep, large-area surveys with ground-based observations is therefore challenging. Although Atacama Large Millimeter Array (ALMA) could in principle  produce deep, high-resolution imaging at 450\,$\mu$m, large surveys would be observationally expensive due to the very limited field of view of $\sim$\,0.02\,arcmin$^{2}$ at this wavelength. In contrast, the SCUBA-2 camera \citep{2013holland} on the JCMT, with a field of view of 
45\,arcmin$^2$ and a beam size of 7.9\,arcsec at 450\,$\mu$m (yielding an approximately 20 times lower confusion limit than SPIRE at 500\,$\mu$m), provides the sensitivity, mapping speed and angular resolution necessary to identify 450-$\mu$m sources and their counterparts over fields of 100s arcmin$^2$. 

Studies of sources selected at 450\,$\mu$m, closer to the peak of the far-infrared emission in systems at the epoch of peak star formation ($z$\,$\sim$\,2--3), have suggested that the 450-$\mu$m-selected population  at mJy flux limits lies at 
lower redshift than those selected at 850\,$\mu$m, with a distribution that peaks at $z$\,$\sim$\,1.5--2.0 \citep{2013casey,2013geach,2013roseboom,2014zavala,2017zavala,2017bourne,2020lim}. 450-$\mu$m-selected sources are also suggested to have higher characteristic dust temperatures than 850-$\mu$m-selected SMGs by $\Delta T_{\rm d}$\,$\simeq$\,10\,K \citep{2013casey,2013roseboom}, although this may be partly due to selection effects. However, the identification of  
differences in the physical properties of  SMGs selected in the different
submillimetre wavebands from such studies have been  limited by their modest sample sizes and also their biases towards brighter sources due to the sensitivity limits at 450\,$\mu$m. Comparison to the 850-$\mu$m population is further complicated by uncertain or incomplete identifications in 
the longer wavelength samples \citep[e.g.][]{2013hodge}, as well as the use of different photometric redshift and SED modelling methods on different samples. 

This study aims to better understand the physical properties of SMGs and in particular the relationship between samples selected at 450- and 850-$\mu$m, by exploiting a very deep 450-$\mu$m imaging survey: the SCUBA-2 Ultra Deep Imaging EAO Survey \citep[STUDIES;][]{2017wang,2020lim} in the Cosmic Evolution Survey \citep[COSMOS;][]{2007scoville} field. 
STUDIES is a multi-year JCMT survey within the CANDELS region ($\sim$300\,arcmin$^2$), which obtained  the deepest single-dish map at 450\,$\mu$m currently available, with a 1-$\sigma$ depth of 0.65\,mJy. The source catalogue and physical properties of the 450-$\mu$m sample are presented in \cite{2020lim}, while the
structural parameters and morphological properties  have been analysed by \citet{2018chang}. 

In this paper, we  compare the properties of  galaxies selected from deep 450-$\mu$m-observations  to those selected from typical 850-$\mu$m surveys.  SCUBA-2 simultaneously maps at 450\,$\mu$m and 850\,$\mu$m, however, the STUDIES 850-$\mu$m map is confusion limited at an 850-$\mu$m flux limit of $\sim$\,2\,mJy and it cannot be used to reliably identify faint 850-$\mu$m sources. Therefore, for our 850-$\mu$m comparison sample we utilise the largest available ALMA-identified 850-$\mu$m-selected SMG sample, from the ALMA/SCUBA-2 Ultra Deep Survey (AS2UDS; \citealt{2019stach}, \citetalias{2020dudz}). 

We revisit the modelling of the UV-to-radio SEDs of the 450-$\mu$m galaxies from STUDIES using {\sc magphys}+photo-$z$ \citep{2008dacunha,2015dacunha,2019battisti},
which was employed by \citetalias{2020dudz} on the AS2UDS 850-$\mu$m sample, thus ensuring that the comparison of the physical properties between the two samples  is free from systematic differences due to the modelling.
With these two large, consistently analysed samples we empirically compare the physical properties of 450-$\mu$m-detected galaxies with the 850-$\mu$m  population. At the observed depths, the two surveys sample down to the ULIRG/LIRG limit (10$^{11-12}$\,$L_\odot$) at $z$\,=\,1--3 and although there is some overlap between these flux-limited samples, as discussed in \citet{2020lim}, the combination and comparison of 450-$\mu$m and 850-$\mu$m surveys provide a more complete view of  luminous far-infrared activity
in the Universe over a wider redshift range than possible with either individual sample.
In particular, we exploit these large samples to construct subsets that are matched in rest-frame wavelength to allow us to quantify the physical differences between an identically selected sample of dusty galaxies at $z$\,$\sim$\,1.5 and $z$\,$\sim$\,3.5.

The paper is structured as follows. In \S~\ref{observations} we give details on the multi-wavelength data which we use to construct the SEDs for our sources and  describe the SED fitting procedure. In \S~\ref{analysis} we present our results, including a comparison of the STUDIES 450-$\mu$m selected galaxies to the 850-$\mu$m selected galaxies from the AS2UDS survey. We discuss the implications of our results in \S~\ref{discussion} and present our conclusions in \S~\ref{conclusions}. We adopt a $\Lambda$CDM cosmology with with $H_0 $\,$=$\,70\,km\,s$^{-1}$\,Mpc$^{-1}$, $\Omega_\Lambda$\,$=$\,0.7, $\Omega_{\rm m} $\,$=$\,0.3. When quoting magnitudes, we use the AB photometric system.

\section{Observations \& SED Fitting} \label{observations}

\subsection{Photometric coverage}

STUDIES is a SCUBA-2 450-$\mu$m imaging survey within the CANDELS region in the COSMOS field. A detailed description of the SCUBA-2 observations and data reduction can be found in \cite{2017wang} and \cite{2020lim}. Briefly, the data from STUDIES, combined with archival data taken by \citet{2013geach} and \citet{2013casey}, yields  the deepest single-dish map currently available at 450\,$\mu$m, reaching a 1-$\sigma$ noise level of 0.65\,mJy. This survey detects 256 sources with a signal-to-noise ratio (SNR) of SNR\,$\geq$\,4 
(of which 126 have SNR\,$\geq$\,5) in an area of 300\,arcmin$^2$.
The confusion-limited 850-$\mu$m map reaches an instrumental noise level of 0.10\,mJy in the deepest regions and has an estimated confusion noise of 0.7\,mJy \citep{2020lim}. The 850-$\mu$m flux densities of the 450-$\mu$m-selected STUDIES sources were obtained from the 850-$\mu$m map at the 450-$\mu$m positions. The source is classed as detected at 850\,$\mu$m if the flux density has SNR\,$\geq$\,5, otherwise it is treated as a limit.

In this section, we provide a brief description of the counterpart identification and multi-wavelength photometric data available for the sample from UV to radio wavelengths, which is then used to model the SEDs of the sources. For a full description of the photometric data and counterpart identification for the STUDIES 450-$\mu$m sample, please refer to \citet{2020lim}.

\subsubsection{Counterpart identification} \label{sec:counterpart}

The identification of optical counterparts for the STUDIES 450-$\mu$m sources is described in \citet{2020lim}. Briefly, the 450-$\mu$m sources were matched with the VLA-COSMOS 3-GHz catalogue \citep{2017smolcic} using a 4\,arcsec search radius (set by the JCMT 450-$\mu$m beam) yielding $\sim$\,1 per cent false positive rate (based on the probability of false matches using the number densities of both catalogues). For the 450-$\mu$m sources above SNR\,$\geq$\,5 this yielded 89/126 counterparts (and 134/256 for SNR\,$\geq$\,4). These radio counterparts were cross matched with the \textit{Spitzer} IRAC catalogue \citep{2007sanders} using a 1\,arcsec search radius with a $\sim$\,3 per cent false positive rate. For the 450-$\mu$m sources that did not have 3-GHz radio counterparts, these were cross matched with the \textit{Spitzer} MIPS 24-$\mu$m catalogue \citep{2007sanders} with a search radius of 4\,arcsec resulting in 27/37 matches for 450-$\mu$m sources with SNR\,$\geq$\,5 (and 76/122 for SNR\,$\geq$\,4). These MIPS counterparts were in turn used to find IRAC counterparts within 2\,arcsec with a $\sim$\,2 per cent false positive rate. The identification rates in different ancillary bands is presented in \citet{2020lim}.
The remaining ten SNR\,$\geq$\,5 450-$\mu$m sources with no radio or MIPS counterparts (and the
46 with SNR\,$\geq$\,4) were matched using a 1\,arcsec matching radius to the catalogue of colour/radio-selected candidate submillimetre counterparts from 
\citet{2019an}.  This catalogue was constructed  using a radio+machine-learning method applied to  a training set comprising ALMA-identified 870-$\mu$m SMGs in the COSMOS and UDS  fields with
the goal of identifying  multi-wavelength counterparts of S2COSMOS  850-$\mu$m single-dish detected sub-millimetre sources. This produced  five identifications for the SNR\,$\geq$\,5 sources (and 15  at SNR\,$\geq$\,4).

For the SNR\,$\geq$\,5 450-$\mu$m sample this process yields reliable counterparts for 121/126 (96 per cent) of the sources, which declines to 207/256 (81 per cent) for those with SNR\,$\geq$\,4. 
As we wish to have a highly-complete and hence unbiased sample, in this paper we analyse the SNR\,$\geq$\,5 450-$\mu$m sources, equivalent to $S_{450}\geq$\,3.25\,mJy, which have almost complete identifications. For this sample of 126 sources: 
89  are located through radio counterparts, 27  are identified through MIPS counterparts and from the remaining ten sources, 
five have counterparts derived from the machine-learning method. In total 109 of the counterparts have IRAC detections. Although some studies have shown that mis-identifications are possible due to the large beam sizes ($\sim$\,15--20\,arcsec FWHM) of single-dish telescopes at long wavelengths \citep[e.g.][]{2013hodge}, this  is much less of an issue for the 450-$\mu$m beam (7.9\,arcsec FWHM) and is further reduced by the SNR\,$\geq$\,5 cut as the centroid position is more precise ($\lesssim$\,1\,arcsec). 

For our sample, we find that, on average, the sources are detected in 19 bands (16--84th percentile range of $N_{\rm det}$\,$=$\,13--22). The detection rate is 70/105 in $B$-band, 87/87 in $z$-band, 115/116 in $H$-band, 107/109 at 4.5\,$\mu$m, 91/121 at 250\,$\mu$m and 43/121 at 850\,$\mu$m.

\subsubsection{Far-infrared to radio observations}

To constrain the SED of each galaxy at radio wavelengths we utilise 1.4-GHz and 3-GHz data from the Very Large Array (VLA)-COSMOS Large Project \citep{2010schinnerer,2017smolcic}. The 3-GHz survey has a noise of 2.3\,$\mu$Jy\,beam$^{-1}$ and an angular resolution of 0.7\,arcsec.  The 1.4\,GHz data is compiled in the COSMOS2015 catalogue \citep{2016laigle} and covers the entire COSMOS field with $\sigma=$\,12\,$\mu$Jy\,beam$^{-1}$ with an angular resolution of 2.5\,arcsec.

At $z$\,$\simeq$\,2, the far-infrared SED of a source with a
characteristic temperature of $T_{\rm d}\sim $\,30\,K is expected to peak at an observed wavelength of $\sim$\,300\,$\mu$m. Hence, to better constrain the shape of the far-infrared SEDs for the galaxies in our sample, and so improve the constraints on the far-infrared luminosities, we include  observations with the Spectral and Photometric Imaging Receiver \citep[SPIRE:][]{2010griffin} and the Photodetector Array Camera and Spectrometer \citep[PACS:][]{2010poglitsch} on the {\it Herschel Space Observatory}. We specifically make use of the 100 and 160\,$\mu$m PACS \citep{2011lutz}, and 250 and 350\,$\mu$m SPIRE observations taken as part of the {\it Herschel} Multi-tiered Extragalactic Survey  (HerMES; \citealt{2012oliver}). 
We adopt the PACS 100- and 160-$\mu$m flux densities from \citet{2011lutz} (as listed in the the COSMOS2015 catalogue),  who presented the observations of the 2\,deg$^{2}$ COSMOS field which reach a 3-$\sigma$ depth of 10.2\,mJy at 160\,$\mu$m.

Due to the coarse resolution of the SPIRE maps ($\sim$\,18\,arcsec and $\sim$\,25\,arcsec FWHM at 250 and 350\,$\mu$m, respectively), we use the method described in \cite{2014swinbank} to deblend these maps and obtain reliable flux densities for our catalogue. The deblending of the SPIRE maps  used positional priors for sources based on the  3-GHz and  24-$\mu$m (see below) catalogues, as well as machine-learning identified SMG counterparts from \cite{2019an} (see \S~\ref{sec:counterpart}). 
The observed flux density distribution is fitted with beam-sized components at the position of a given source in the prior catalogue using a Monte Carlo algorithm. The method is first applied to the 250-$\mu$m data, then to avoid ``over-blending'' only sources that are detected at $>$\,2-$\sigma$ at 250\,$\mu$m are propagated to the prior list for the 350-$\mu$m deblending. The uncertainties on the flux densities (and limits) are found by attempting
to recover model sources injected into the maps (see \citealt{2014swinbank} for details), and yield typical 3-$\sigma$ detection limits of 7.0 and 8.0\,mJy at 250 and 350\,$\mu$m, respectively\footnote{Comparison of our measurements to the deblended {\it Herschel} sources from \citet{2018jin} showed agreement within the quoted errors with abs(S-S$_{\rm Jin+2018}$)/S$_{\rm err}$ of 1.25 and 1.15 at 250\,$\mu$m and 350\,$\mu$m, respectively.}.

\subsubsection{Optical to near-/mid-infrared observations}

To model the stellar SEDs of the counterparts to our 450-$\mu$m sources, we require the photometry in the optical/infrared bands. For the $u^*Bgrizy$ bands we adopt the photometry from COSMOS2015 \citep{2016laigle} catalogue. The $u^*$-band data is from the Canada-France-Hawaii Telescope \citep[CFHT/MegaCam;][]{2003boulade} and covers the entire COSMOS field with a 5-$\sigma$ depth of $u^*$\,$\simeq$\,26.5 \citep{2009ilbert}. The $B$-band imaging was taken with Subaru Suprime-Cam as part of COSMOS-20 survey \citep{2007taniguchi} and has a 5-$\sigma$ depth of $B$\,$=$\,27.2 in a 2\,arcsec diameter aperture. Images in the $grizy$-bands are taken from the second data release (DR2) of the Hyper-SuprimeCam (HSC) Subaru Strategic Program \citep[SSP;][]{2019aihara}.  The nominal  5-$\sigma$ depths are $g$\,$=$\,27.3, $r$\,$=$\,26.9, $i$\,$=$\,26.7, $z$\,$=$\,26.3 and $y$\,$=$\,25.3 in 2\,arcsec diameter apertures.

In addition we employ $YJHK_{\rm s}$ imaging  from the fourth data release (DR4) of the UltraVISTA survey \citep{2012mccracken}. In an equivalent manner to \citet{2020simpson}, we measure 2\,arcsec diameter aperture photometry at the positions of each SMG in each band. The uncertainty on the derived flux densities is estimated in a 1\,$\times$\,1\,arcmin$^2$ region centred on the position of each SMG. Finally, we convert the derived flux densities to a total flux density by applying an aperture correction of a factor of 1.80, 1.74, 1.52 and 1.46 for the $YJHK_{\rm s}$ bands, respectively. This is done by comparing the DR4 photometry to the UltraVISTA DR2 photometry from COSMOS2015, for those SMGs with a counterpart in the catalogue. 

For the near-infrared photometry, we employ the {\it Spitzer} IRAC data from \cite{2007sanders}. IRAC 3.6-, 4.5-, 5.8- and 8.0-$\mu$m imaging was obtained as part of the S-COSMOS  survey, which covers the entire COSMOS field and has an angular resolution of 1.7\,arcsec at 3.6\,$\mu$m.  The 24-$\mu$m catalogue was generated by \citet{2020lim}, who used the S-COSMOS 24\,$\mu$m image \citep{2007sanders}. The catalogue has a 3.5-$\sigma$ limit of 57\,$\mu$Jy. 

We correct the $u^*$-band to IRAC 8.0-$\mu$m photometry of each source for Galactic extinction based on its sky position, the extinction maps of \cite{2011schlafly}, and the extinction curve of \cite{1999fitzpatrick}, assuming a reddening law with $R_V$\,$=$\,3.1. For each filter, the correction is determined by convolving the filter response with the scaled extinction curve. 

%
%
\begin{figure*}
  \includegraphics[width=\textwidth]{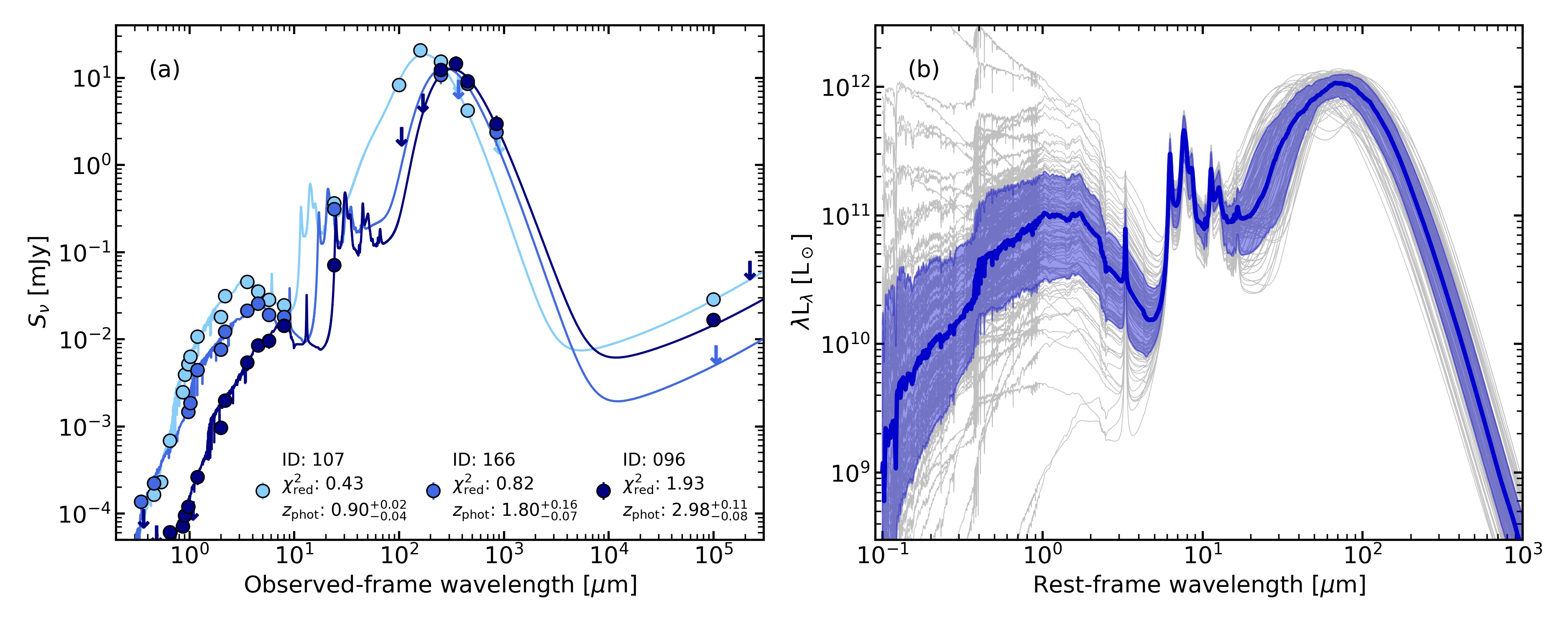}
  \caption{
  {\bf (a)} Example observed-frame optical-to-radio SEDs for three STUDIES sources selected from the $\sim$\,16th, $\sim$\,50th and $\sim$\,84th percentiles of the STUDIES 450-$\mu$m redshift distribution. The solid lines show the SEDs at the peak redshift of the corresponding best-fitting model from {\sc magphys}+photo-$z$. The arrows indicate any  upper limits in the photometric observations. We see that the SEDs are well constrained at these redshifts, since the 450-$\mu$m sample is detected on average in 19 bands, with a 16--84th percentile range of 13--22 bands.
  {\bf (b)} Best-fitting SEDs of all 121 STUDIES SMGs normalised to the median far-infrared luminosity of the sample, $L_{\rm IR}$\,$=$\,1.6\,$\times$\,10$^{12} $\,L$_\odot$. We also overlay the median composite SED and indicate the 16--84th percentile region as the shaded area. We observe that the 450-$\mu$m SMGs display a wide variety of optical luminosities and colours spanning an order of magnitude at rest-frame $K$-band and around twice that at shorter wavelengths highlighting the difficulty to produce complete samples of highly star-forming galaxies using UV/optical observations alone.}
  \label{fig:composite1}
\end{figure*}

\subsection{SED fitting model}

To derive the physical properties of the STUDIES SMGs we employ {\sc magphys}+photo-$z$ \citep{2015dacunha,2019battisti} to model the SEDs from optical to radio wavelengths, using the available photometry in 24 bands. The attenuation of the stellar emission by dust in the UV/optical and near-infrared and the consequent re-radiation in the far-infrared is coupled via an energy balance technique. This model allows us to constrain the physical parameters of the galaxies, as well as providing a consistent methodology to that applied to the large ALMA-identified 850-$\mu$m SMG sample from the AS2UDS survey by \citetalias{2020dudz}. Thus, the physical properties of the two samples can be investigated for any differences arising from the different wavelength selection.

{\sc magphys}+photo-$z$ uses stellar population models from \citet{2003bruzual&charlot} and a \citet{2003chabrier} IMF. Star-formation histories (SFH) are modelled as continuous delayed exponential functions
\citep{2010lee} with superimposed bursts. Dust attenuation is modelled by a two-component model of \citet{2000charlot&fall}, combining the effective attenuation from dust in stellar birth clouds and diffuse interstellar medium, parameterised by the reddening in the $V$-band.
The star-formation rate is calculated using the best-fitting model star-formation history, after accounting for dust attenuation and is defined as the average of the SFH over the last 10\,Myr.
The far-infrared emission from dust in {\sc magphys}+photo-$z$ is determined self-consistently from the dust attenuated stellar emission.
The far-infrared luminosity is measured by integrating the SED in the rest-frame between 8--1000\,$\mu$m and is calculated through the sum of the birth cloud and ISM luminosities, including contributions from the polycyclic aromatic hydrocarbons, and mid-infrared continuum from hot, warm and cold dust in thermal equilibrium. The dust mass is calculated fitting a two-component modified blackbody with emissivity index, $\beta$, fixed at 1.5 for the warm components and 2.0 for the cold components. 
We note that different assumptions in model emissivity index, dust opacity and dust mass absorption coefficient will impact the dust mass measurements, therefore care must be taken when interpreting and comparing results \citep[see][]{2014casey}.
{\sc magphys}+photo-$z$ estimates a characteristic dust temperature using five free parameters that combine the contribution from the warm (birth clouds) and cold (diffuse ISM) components.
A full description of {\sc magphys}+photo-$z$ code and parameter derivation can be found in \citet{2015dacunha} and \citet{2019battisti}.

To fit the observed multi-wavelength photometry of each galaxy, for each star-formation history {\sc magphys}+photo-$z$ creates a library of SEDs at random redshifts, resulting in $\sim$\,300,000 templates. The best-fitting SED is selected using a $\chi^2$ test returning the photometric redshift, best-fitting parameters and their probability distributions. The uncertainties for each parameter are taken as 16--84th percentile of the probability distribution. For our analysis, we used the updated {\sc magphys}+photo-$z$ code from \citet{2015dacunha} and \citet{2019battisti}, which is optimised for high redshift star-forming galaxies with extended prior distributions for dust optical depths and star-formation histories. The code also includes a parameterisation to reproduce the intergalactic medium absorption of UV photons. 

We note that the far-infrared SEDs of our SMGs are covered by at most six photometric bands (see Fig.~\ref{fig:composite1}a), with weaker constraints  near the peak of the SEDs, hence to provide a robust estimate of dust temperature, we adopt a conservative approach and fit a single modified blackbody to the available {\it Herschel}  100-, 160-, 250- and 350-$\mu$m
photometry and the SCUBA-2 flux densities at 450\,$\mu$m and 850\,$\mu$m. This approach allows for a simple comparison with similar fits to other 450-$\mu$m, 850-$\mu$m and ALMA samples. 
We estimate the characteristic dust temperature using a modified black body of the form: 
\begin{equation}
    S_{\nu_{\mathrm{obs}}} \propto (1 - e^{-\tau_{\mathrm{rest}}}) \times B(\nu_{\mathrm{rest}}, T),
    \label{eq:bb}
\end{equation}
where $B(\nu_{\mathrm{rest}}, T)$ is the Planck function, $\tau_{\mathrm{rest}}$ is the frequency-dependent optical depth of the dust of the form $\tau_{\mathrm{rest}}$\,$=$\,$\Big(\frac{\nu_{\rm rest}}{\nu_0}\Big)^\beta$, $\nu_0$ is the frequency at which optical depth is equal to one and $\beta$ is the dust emissivity index. We adopt $\beta$\,$=$\,1.8, as used in previous SMG studies and consistent with that estimated for local star-forming galaxies \citep{2011planck,2013clemens,2013smith}. For the purpose
of calculating canonical values we 
adopt an optically thin prescription to describe the region from which the dust emission originates, thus $\nu_0 \gg \nu_{\rm rest}$ and Eq.~\ref{eq:bb} simplifies to, $S_{\nu_{\mathrm{obs}}} \propto \nu_{\mathrm{rest}}^\beta \times B(\nu_{\mathrm{rest}}, T)$. However, we stress that it is likely 
that the dust emission from the sources in our sample is not optically 
thin in the far-infrared and  $T^{\rm MBB}_{\rm d}$ and $M_{\rm d}$ (through strong dependence on $T^{\rm MBB}_{\rm d}$) are affected by the dust opacity assumptions. If we instead adopted a modified black body with an opacity term (Eq.~\ref{eq:bb}) systematic offsets arise, with lower $\nu_{0}$ leading to higher  $T^{\rm MBB}_{\rm d}$ and thus $M_{\rm d}$ \citep[see][]{2014casey}. 

We note that there is a systematic offset of $L^{\rm MBB}_{\rm IR}$/$L_{\rm IR}=$\,0.85 between the far-infrared luminosity calculated from a modified black body fit and {\sc magphys}+photo-$z$ fit, which includes the emission on the Wein side of the SED. Thus, we apply this factor to any subsequent modified black body luminosities (see \S~\ref{sec:farir}) to homogenise the two fitting methods. The systematic offset between the dust masses retrieved using these two different methods is $\lesssim$\,10 per cent, which is within the uncertainty of the {\sc magphys}+photo-$z$ dust mass values.

 We run {\sc magphys}+photo-$z$ on the available photometry of our 121 SNR\,$\geq$\,5 450-$\mu$m sources to obtain the best-fitting model SEDs. To correct for calibration differences between our multiwavelength photometry and the stellar template libraries used in {\sc magphys}+photo-$z$,  we follow \cite{2006feldmann} and iteratively
determine the systematic offsets between the observed photometry and the predicted photometry from the best-fitting model SEDs. We estimate the median difference at each band for our whole sample and use these values to adjust the zero-points of the filters from the $u^*$ band to IRAC 8.0\,$\mu$m. 
The applied fractional offsets to the flux densities are: $u^*$ ($-$0.165); $B$ ($-$0.023); $g$ ($+$0.019); $r$ ($-$0.009); $i$ ($+$0.062); $z$ ($+$0.103); $y$ ($-$0.041); $Y$ ($-$0.024); $J$ ($-$0.030); $H$ ($+$0.036); $K_{\rm s}$ ($+$0.061); 3.6\,$\mu$m ($-$0.037); 4.5\,$\mu$m ($-$0.108); 5.8\,$\mu$m ($-$0.135); 8.0\,$\mu$m ($-$0.233). Given the absence of an AGN component in the {\sc magphys}+photo-$z$ modelling and the relative paucity of constraints at far-infrared and sub-millimetre wavelengths, we do not apply any offsets at wavelengths beyond 8\,$\mu$m. We then re-run {\sc magphys}+photo-$z$ with the adjusted photometry to obtain the best-fitting SEDs (all SEDs are shown in Fig.~\ref{fig:composite1}b), redshift and physical properties for the 450-$\mu$m sources. In case of limits in any given band, we adopt values of 0$\pm$3$\sigma$ in the optical-to-8\,$\mu$m and 1.5$\sigma \pm$1$\sigma$ in the 8\,$\mu$m-to-radio bands. Three example best-fitting SEDs of sources residing at the 16th, 50th and 84th percentiles of the STUDIES 450-$\mu$m redshift distribution (discussed in \S~\ref{sec:redshift}) are shown in Fig.~\ref{fig:composite1}b.

A detailed description of testing of the {\sc magphys}+photo-$z$ code on large samples of high-redshift observed and simulated SMGs can be found in \citetalias{2020dudz}. The study applied {\sc magphys}+photo-$z$ to a sample of $\sim$\,7,000 galaxies with spectroscopic redshifts in the UDS field to test the precision of photometric redshifts. For the 44 850-$\mu$m SMGs with spectroscopic redshifts, they determined a fractional offset of $\Delta z/(1+z_{\rm spec})=$\,$-$0.02\,$\pm$\,0.03, with a 1-$\sigma$ range of $-$0.16--0.10, which is comparable to the accuracy found in \cite{2019battisti}. From our 450-$\mu$m sample, 32 sources have spectroscopic redshifts and we find a similar fractional offset of $\Delta z/(1+z_{\rm spec})=-$0.03\,$\pm$\,0.04, with a 1-$\sigma$ range of $-$0.27--0.11. We also test for systematic differences between the {\sc magphys}+photo-$z$ redshifts determined in this work with those used by \cite{2020lim}. The latter comprises a mix of spectroscopic redshifts,  photometric redshifts from \citet{2016laigle} and cruder redshift estimates from fitting an SMG template SED.  We determine a  fractional redshift offset for the 450-$\mu$m SNR\,$\geq$\,5 sample of $(z_{\rm MAGPHYS}-z_{\rm Lim})/(1+z_{\rm MAGPHYS}) =-$0.007\,$\pm$\,0.013. We conclude that on average the redshifts derived here are consistent with previous estimates for this sample.

We note that, for consistency, we use the {\sc magphys}+photo-$z$ derived photometric redshifts for all sources in the analysis in this paper. To test the effect of photometric redshift estimates on the predicted physical properties (which are presented in \S~\ref{properties}) we run {\sc magphys}+photo-$z$ on the 32 450-$\mu$m sources at their fixed spectroscopic redshifts. We calculate the fractional difference ($X_{\rm spec}-X_{\rm phot})/X_{\rm phot}$, where $X$ is a given physical parameter, between the physical parameter values derived at the spectroscopic and photometric redshifts. We find that the typical systematic offset for any given physical property (SFR, $L_{\rm FIR}$, $M_\ast$, $M_{\rm d}$, $A_{\rm V}$) is $\sim$\,10 per cent, which is within the typical uncertainties. Therefore, the redshift uncertainty effect on any given parameter is captured within it's error range.

%
%
\begin{figure*}
  \includegraphics[width=\textwidth]{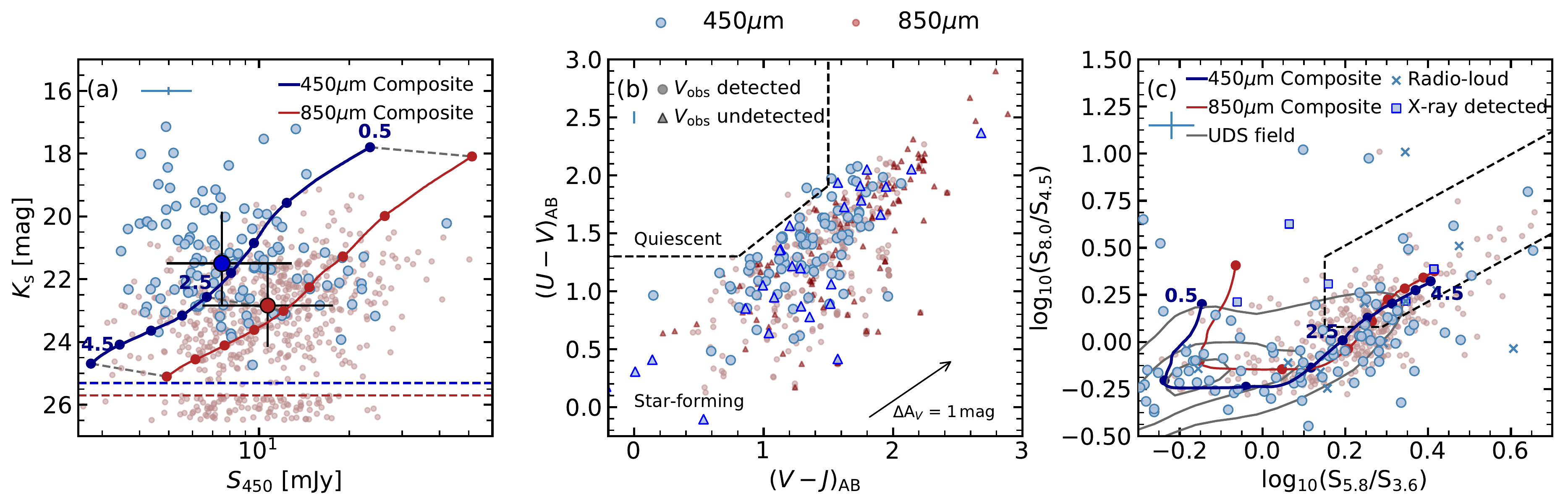}
  \caption{{\bf (a)} The distribution of $K_{\rm s}$-band magnitude versus 450-$\mu$m flux density for our SNR\,$\geq$\,5.0 450-$\mu$m sample. For comparison, we also overlay the wide-field 850-$\mu$m SMG sample from \protect\citetalias{2020dudz}, for which the 450-$\mu$m flux density is estimated from the {\sc magphys}+photo-$z$ best-fitting SEDs. We indicate the median for each sample as a large circle in the respective colour, with the 16--84th percentile range shown as the black error bar. 450-$\mu$m-selected sources cover a similar range in 450-$\mu$m flux density to the 850-$\mu$m population, but, on average, have brighter $K_{\rm s}$-band magnitudes, likely reflecting their lower redshifts, higher stellar masses and/or lower dust attenuation. We overlay the tracks of the composite SEDs in $\Delta z$\,$=$\,0.5 increments. The grey dashed lines show the difference in SEDs at the same redshift and indicates that both samples have similar $K_{\rm s}$-band magnitudes at a given redshift, suggesting that redshift is the main driver of the differences in $K_{\rm s}$-band brightness between the two samples. The limiting $K_{\rm s}$-band magnitude for each sample is indicated as the dashed line in the respective colour. The median error on any individual source is shown in the top left in light blue.
  {\bf (b)} Rest-frame $(U-V)$ versus $(V-J)$ colour-colour diagram for 450-$\mu$m- and 850-$\mu$m-selected sources detected in the observed-frame $J$ and 4.5-$\mu$m bands. We indicate those detected (circles) and undetected (triangles) in the observed $V$-band. We overlay the selection criteria for star-forming galaxies from \protect\citet{2012whitaker}. 
  We observe that the bulk of the STUDIES and AS2UDS sources have colours consistent with them being star forming, but potentially $\sim$\,5 per cent have colours or limits which could place them in the ``Quiescent" classification even though they are likely to be strongly star forming. The rest-frame values were obtained from the {\sc magphys}+photo-$z$ best-fit SEDs. The error due to redshift uncertainty on each source is shown in the top right. The reddening vector for one magnitude of extinction in the $V$-band is shown in the bottom right of the panel.
  {\bf (c)} IRAC colour-colour diagram for the 450-$\mu$m sources. We indicate radio-loud sources and those with an X-ray-detected counterpart as identified by \citet{2020lim}. The dashed lines indicate the IRAC selection criteria for AGN at $z$\,$\leq$\,2.5 from \protect\cite{2012donley}. For comparison, we overlay the AS2UDS sources with $z\leq$\,3 and the tracks of the composite SED of 850-$\mu$m AS2UDS SMGs and 450-$\mu$m STUDIES SMGs at $z$\,$=$\,0.5--4.5 redshifts in $\Delta z$\,$=$\,0.5 increments. We also overlay a contour of the density of $K$-band selected UDS field galaxies, which shows that both SMG samples have redder IRAC colours than the field population. 850-$\mu$m sources have colours which cluster in a region that matches their redshifted template at $z$\,$\sim$\,1.5--4 and are typically redder than the 450-$\mu$m selected population which have colours consistent with their template in a broad range at $z$\,$\sim$\,1--3.}
  \label{fig:obs}
\end{figure*}

\section{Analysis \& Results} \label{analysis}

In this section, we investigate the broad photometric properties and the derived physical properties of the 450-$\mu$m sample based on our {\sc magphys}+photo-$z$ analysis of their SEDs. We compare the results of the 450-$\mu$m-selected sample to an 850-$\mu$m-selected sample, AS2UDS \citep{2019stach}, which has been analysed in a consistent manner by fitting {\sc magphys}+photo-$z$ to the available photometry in 22 bands \citepalias{2020dudz}.

AS2UDS is a follow-up survey of sources
detected in the SCUBA-2 Cosmology Legacy Survey (S2CLS, \citealt{2017geach}) 850-$\mu$m map of the $\sim$\,0.9\,deg$^2$ UKIDSS UDS field and provides a large homogeneously-selected sample of ALMA-identified SMGs. The parent SCUBA-2 sample reaches a 4-$\sigma$ limit of $S_{850}$\,$=$\,3.6\,mJy, with the 707 ALMA detected counterparts having 870-$\mu$m flux densities spanning $S_{870}$\,$=$\,0.6--13\,mJy (see \citealt{2019stach} and \citetalias{2020dudz}, for more details). Throughout the paper, we will refer to this as the 850-$\mu$m sample (we note that this ALMA selection formally corresponds to 870$\mu$m, which is the wavelength used in the analysis). 

\subsection{Photometric properties of 450-\texorpdfstring{$\mu$}m sources} \label{sec:phot props}

Before we discuss the physical properties of the STUDIES 450-$\mu$m SMGs in detail, we first investigate the observed and rest-frame optical and infrared colour properties of the sample. Throughout the paper we compare the 450-$\mu$m-selected population to the AS2UDS 850-$\mu$m-selected sample, hence we start by comparing the features of the near- and far-infrared photometry of the two samples to assess their broad properties and where they fall in two commonly used photometric classifications. By assessing the best-fitting {\sc magphys}+photo-$z$ SEDs, we find that 14 per cent of 450-$\mu$m-selected SMGs are expected to be brighter than the UDS SCUBA-2 850-$\mu$m flux density limit of $S_{850}$\,$=$\,3.6\,mJy at 850\,$\mu$m. Similarly, if we look at the 850-$\mu$m sources from AS2UDS, we find that 98 per cent are expected to be brighter than 3.25\,mJy at 450\,$\mu$m (the STUDIES limit). This suggests that the deep, but relatively narrow-field, STUDIES 450-$\mu$m survey is probing to lower dust masses than the shallower, wide-field AS2UDS 850-$\mu$m survey.

In Fig.~\ref{fig:obs}a we show the distribution of $K_{\rm s}$-band magnitude with  450-$\mu$m flux density for
our 450-$\mu$m sources and the full 850-$\mu$m sample. For 850-$\mu$m sources, we show the predicted 450-$\mu$m flux density based on the model SEDs for each source. The two populations have comparable 450-$\mu$m flux densities, with the 850-$\mu$m sample having a slightly higher predicted median. The 450-$\mu$m sample has considerably brighter $K_s$-band magnitudes, with the median ($K_{\rm s}=$\,21.5$\pm$0.2\,mag) being comparable to the 16th percentile value ($K_{\rm s}=$\,21.52$\pm$0.06\,mag) of the 850-$\mu$m sample. We also note that \citetalias{2020dudz} found that 17 per cent of the 850-$\mu$m SMGs are undetected in the $K_{\rm s}$-band, while the upper limit of $K_{\rm s}$-band non-detections in the 450-$\mu$m sample is $<$\,5/126 ($<$\,4 per cent), corresponding to only those sources lacking counterparts. The results indicate that the 450-$\mu$m population is potentially at lower redshift, has higher stellar masses and/or has lower dust attenuation.

%
%
\begin{figure*}
  \includegraphics[width=\textwidth]{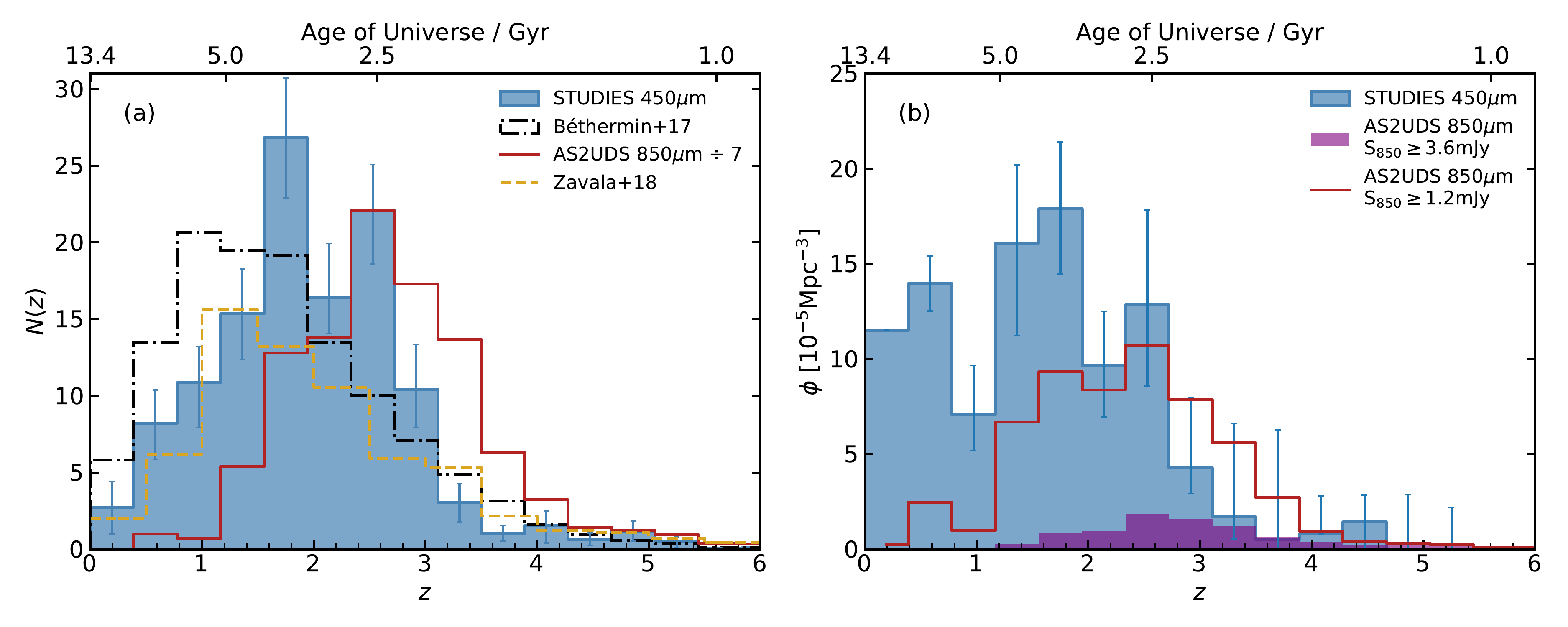}
  \caption{{\bf (a)} Stacked likelihood redshift distribution for the 121 sources selected at 450\,$\mu$m  with a median redshift of $z$\,$=$\,1.85\,$\pm$\,0.14 and 9 per cent of the sample at lying at $z$\,$\geq$\,3. For comparison, we overlay the stacked likelihood distribution of the sample of 65 450-$\mu$m detected sources from \citet{2018zavala}, which has a similar median redshift, and the $\sim$\,700 850-$\mu$m selected SMGs from the AS2UDS survey (scaled down in normalisation by a factor of seven) which have a median redshift of $z$\,$=$\,2.61\,$\pm$\,0.08. We also show the predicted redshift distribution for a sample with $S_{450}$\,$>$\,3.25\,mJy from the galaxy evolution model by \citet{2017bethermin}, normalised to the number of sources in our 450-$\mu$m sample, which peaks at lower redshift than our sample. Equally, the redshift
  distribution of the 450-$\mu$m sources peaks at  lower redshifts than 850-$\mu$m sources, which can be explained by the luminosity function and luminosity-redshift relation for these selections. The errors shown on the 450-$\mu$m histogram are estimated from the bootstrap resampling of the redshift likelihood distributions.
  {\bf (b)} Variation in the space density of our 450-$\mu$m-selected population with redshift, showing that the space density of 450-$\mu$m-selected sources decreases with increasing redshift. For comparison, we overlay the redshift distribution for a bright $S_{850}$\,$\geq$\,3.6\,mJy  subset of the 850-$\mu$m sample (this flux density corresponds to the limit of the parent S2CLS survey) after  correcting for the modest incompleteness above this flux density limit using number counts from \citet{2017geach}. To identify an 850-$\mu$m population with more comparable far-infrared luminosities
  to our 450-$\mu$m sample, we estimate the typical 850-$\mu$m flux density for the 450-$\mu$m sample of $S_{850}$\,$\sim$\,1.2\,mJy, using the median 450-$\mu$m flux at $z$\,$\sim$\,2.5 and the composite 450-$\mu$m SED. We therefore also overlay a subset of 850-$\mu$m SMGs with $S_{850}$\,$\gtrsim$\,1.2\,mJy, where the number density has been  corrected for incompleteness in the 850-$\mu$m sample below $S_{850}$\,$\sim$\,4\,mJy, adopting the slope of the number counts from \citet{2018hatsukade}. For this subset, we see that the 450-$\mu$m-selected sources have a similar space density to an 850-$\mu$m population with a comparable far-infrared luminosity at $z$\,$\sim$\,2--3.}
  \label{fig:pz}
\end{figure*}

The rest-frame optical/near-infrared colours of galaxies have long been employed to classify high-redshift galaxy populations \cite[e.g.,][]{1993smail}.
These methods primarily rely on contrasting the colour measured from a combination of photometric bands that span the Balmer/4000\AA\  break in the galaxy rest-frame to the colour in a pair of bands at longer wavelengths, to
attempt to differentiate between quiescent and star-forming SEDs. One frequently used combination
of passbands is rest-frame $UVJ$, where redder  rest-frame $(U-V)$ colours at similar rest-frame $(V-J)$ colours indicate more quiescent populations. We show the $UVJ$ classification scheme from \citet{2012whitaker}  in Fig.~\ref{fig:obs}b. We plot the rest-frame $UVJ$ colours of the 450-$\mu$m and 850-$\mu$m populations using values from the {\sc magphys}+photo-$z$ best-fitting SED of each source and, to ensure that there are meaningful
constraints on the shape of the SEDs, we require that the sources are detected
in the observed $J$ and 4.5\,$\mu$m bands (roughly corresponding to rest-frame $V$ and $J$ at the typical redshifts). We observe that the bulk of the 450-$\mu$m and 850-$\mu$m populations have $UVJ$ colours consistent with them being star-forming sources and that both populations show a spread which is aligned with the reddening vector. As shown in Fig.~\ref{fig:obs}b, around 5 per cent of the 450-$\mu$m sources have limits (and two sources have colours, although these lie near the classification boundaries) which could place them in the ``quiescent" classification even though they are likely to be strongly star forming. We note that red $(U-V)$ and blue $(V-J)$ rest-frame colours can be mimicked by dusty star-forming galaxies at high redshift \citep{2016chen}.

At $z$\,$\gtrsim$\,1--2 the rest-frame $H$-band, which samples the stellar ``bump" at 1.6\,$\mu$m due to the H$^-$ opacity minimum, is redshifted into the IRAC bands at $\gtrsim$\,3.6\,$\mu$m. Thus, IRAC colours suggesting a peak at these wavelengths can crudely indicate the redshift of a source \citep{2002sawicki} and can also be used to distinguish between star-forming galaxies and AGN, which have power-law emission in the mid-infrared. Across the 450-$\mu$m sample there are 100/109 SMGs with detections in all four IRAC bands, which are shown in Fig.~\ref{fig:obs}c. We consider the \cite{2012donley} selection criteria, which is used to identify AGN at $z$\,$\lesssim$\,2.5. There are 19/109 (17 per cent) sources with colours suggesting power-law spectra (7 at $z$\,$<$\,2.5) and thus consistent with the presence of an AGN, with four radio-loud sources (from \citealt{2020lim}, who applied a redshift-dependent threshold in radio excess) and three detected in the X-rays (see \citealt{2020lim}). We overlay the track of the composite SED of the 450-$\mu$m sample (further discussed in \S\ref{properties}) as a function of redshift, which demonstrates that IRAC-colours indicative of AGN are degenerate with those expected for dusty star-forming galaxies at $z$\,$\gtrsim$\,2--3, where many members of this population lie. This is simply a result of the fact that above $z$\,$\sim$\,3 the 1.6-$\mu$m stellar bump moves into the reddest IRAC band at 8.0\,$\mu$m and thus the IRAC colours of dusty star-forming SMGs mimic an AGN-like power-law behaviour \citep[e.g.][]{2010wang}, and hence this colour selection cannot reliably classify these high-redshift sources.  

For comparison to the 450-$\mu$m population, we also show in Fig.~\ref{fig:obs}c the IRAC colours of 850-$\mu$m sources with $z$\,$\leq$\,3, as well as the track of the 850-$\mu$m AS2UDS composite SED as a function of redshift. We note that the composite SED of the 450-$\mu$m sample has bluer IRAC colours at $z$\,$\lesssim$\,2 compared to the 850-$\mu$m composite (see also \S~\ref{sec:optical}). However, 
at $z$\,$\gtrsim$\,2.5 the colours of both populations are comparable. The 450-$\mu$m population clusters around its composite SED track at colours corresponding to $z$\,$\sim$\,1--3, while the 850-$\mu$m population shows a distribution with colours matching the corresponding composite SED at $z$\,$\sim$\,1.5--4.
We stress that both 450-$\mu$m and 850-$\mu$m-selected populations have much redder IRAC colours, on average, than the $K$-selected field population (UKIDSS UDS; Almaini et al.\ in prep.), likely as a result of their higher dust attenuation and typically higher redshifts.

\subsection{Redshift distribution} \label{sec:redshift}

The redshift distribution is a fundamental quantity providing constraints on formation models for the given population and is also essential for reliable derivation of their intrinsic properties and evolutionary trends. To derive a photometric redshift distribution reflecting the uncertainties in any individual SED fit (and hence the quality of the fitting), we stack the individual redshift likelihood distributions from our {\sc magphys}+photo-$z$ analysis for all of the 450-$\mu$m SMGs and show this in Fig.~\ref{fig:pz}a. For the full sample of 121 (SNR\,$\geq$\,5) 450-$\mu$m SMGs, we measure a median redshift of $z$\,$=$\,1.85\,$\pm$\,0.12. The quoted uncertainty is the combination of the systematic uncertainty from the comparison of {\sc magphys}+photo-$z$ redshifts to spectroscopic redshifts of 6,719 $K$-band galaxies in the UDS \citepalias{2020dudz} and the bootstrap error on the stacked redshift distribution. The SNR\,$\geq$\,5 450-$\mu$m SMG population, which is brighter than $S_{450}$\,$\geq$\,3.25\,mJy, shows a peaked but broad redshift distribution with a 16--84th percentile redshift range of $z$\,$=$\,1.0--2.7 and only 9 per cent of sources lying at $z$\,$\geq$\,3 (Fig.~\ref{fig:pz}a).

Our median redshift is comparable to that derived for a sample of 64 450-$\mu$m-selected galaxies, with $S_{450}$\,$\gtrsim$\,3.6\,mJy, from the S2CLS Extended Groth Strip field by \citet{2018zavala}, 
who found a median redshift of $z$\,$=$\,1.66$\pm$0.18, which is within 1.5\,$\sigma$ of our result (before considering excess variance). Similarly, the 450-$\mu$m SCUBA-2 survey of 78 SMGs above a flux density $S_{450}$\,$\simeq$\,15\,mJy (3.6\,$\sigma$ significance) in the COSMOS field by \citet{2013casey} yielded a median redshift of $z$\,$=$\,1.95\,$\pm$\,0.19, also agreeing with our result.

One of the aims of this paper is to study the relationship between the 450-$\mu$m and 850-$\mu$m selected populations, thus we investigate how the selection wavelength affects the redshift distribution at our observed depths. The 850-$\mu$m sample from AS2UDS has a median redshift of $z$\,$=$\,2.61\,$\pm$\,0.08, with a 16--84th percentile range of $z$\,$=$\,1.8--3.4 (see Fig.~\ref{fig:pz}a). We note that the subset with $S_{850}$\,$\geq$\,3.6\,mJy (the completeness limit of the SCUBA-2 parent survey) has a very similar shape with a slightly higher median redshift of $z$\,$=$\,2.78\,$\pm$\,0.09. The median redshifts of the 850-$\mu$m and 450-$\mu$m samples are  different at $\sim$\,5.5\,$\sigma$ significance. We also compare the distributions using a two sample Kolmogorov-Smirnov (K-S) test and find a probability of $P$\,$=$\,2\,$\times$\,10$^{-13}$, indicating that the two distributions are significantly different. This result is contrary to the findings of \citet{2013casey}, who suggest that the 450-$\mu$m and 850-$\mu$m populations occupy a similar redshift range at these flux density limits. The disagreement is due to their 850-$\mu$m sample having a considerably lower median redshift of $z$\,$=$\,2.16\,$\pm$\,0.11, most likely a result of incompleteness in their identifications compared to the ALMA-located AS2UDS survey.

In Fig.~\ref{fig:pz}a, we also compare to a simple model of galaxy evolution based on the observed evolution of the stellar mass function, the main-sequence of star-forming galaxies, and the SEDs by \citet{2017bethermin}, who address the selection effects on the redshift distribution (we discuss this further in \S~\ref{sec:farir}). Their predicted distribution is roughly similar to that of the 450-$\mu$m population, but with a lower median redshift of $z$\,$=$\,1.58\,$\pm$\,0.01 for $S_{450}$\,$>$\,3.25\,mJy sources. This 2-$\sigma$ difference suggests that the luminosity function evolution might have to be stronger than that adopted in their model, in order to produce more sources at higher redshift and thus match the observed 450-$\mu$m redshift distribution.

To compare the 450 and 850\,$\mu$m selections in more detail, we take advantage of our well-defined and almost effectively complete redshift distribution to investigate the space density of the 450-$\mu$m-selected population. We highlight that STUDIES 450-$\mu$m survey area of $\sim$\,300\,arcmin$^2$ is $\sim$\,10 times smaller than the AS2UDS 850-$\mu$m survey area of $\sim$\,3200\,arcmin$^2$. Since the depth of the 450-$\mu$m map varies due to the map coverage, each source has a different effective survey area. Thus, we estimate the survey area for each of the sources by calculating the area within the SCUBA-2 map within which each SMG would be detectable at SNR\,$\gtrsim$\,5, given their 450-$\mu$m flux density, using the 450-$\mu$m RMS map from \citet{2020lim}. The space density is then calculated in redshift bins using the median redshift estimates of each of the SMGs. We estimate the uncertainties by resampling the redshift probability distribution of each source 500 times. The corresponding space density as a function of redshift is shown in Fig.~\ref{fig:pz}b. There is a decrease in space density for the 450-$\mu$m-selected population with increasing redshift, which is particularly marked above $z$\,$\gtrsim$\,2.5. 

For comparison, we calculate the space density evolution for the 850-$\mu$m sample with $S_{850}$\,$\geq$\,3.6\,mJy. This is the flux density limit of the AS2UDS's parent survey - the UDS field of S2CLS \citep{2017geach} that covers an area of 0.96\,deg$^2$ \citep[see][]{2019stach}. This flux limit corresponds to $L_{\rm FIR}$\,$\gtrsim$\,2\,$\times$\,10$^{12}$\,L$_\odot$ (meaning these
sources are all ULIRGs) for typical dust temperatures, compared to $L_{\rm FIR}$\,$\gtrsim$\,0.5--1\,$\times$\,10$^{12}$\,L$_\odot$ for our 450-$\mu$m sample, corresponding to the bright-end of the LIRG population. We correct for incompleteness in the SCUBA-2 850-$\mu$m sample in the UDS field following \citet{2017geach}. As seen in Fig.~\ref{fig:pz}b, the space density for the $S_{850}$\,$\geq$\,3.6\,mJy subset of the AS2UDS SMGs is significantly lower than the 450-$\mu$m population, however, this is primarily due to the different flux density and luminosity limits of the two studies. To better compare the two populations, we calculate the median $S_{450}/S_{850}$ ratio using the composite 450-$\mu$m SED (further discussed in \S~\ref{properties}) and find that an average 450-$\mu$m source at $z$\,$\sim$\,2.5 is expected to have an 850-$\mu$m flux density of $S_{850}$\,$\sim$\,1.2\,mJy. Thus, we select all of the 850-$\mu$m SMGs above this flux density limit and correct for the survey completeness using the ALMA 1.13-mm number counts in the GOODS-S field from \citet{2018hatsukade}. 
The corresponding space density for the complete $S_{850}$\,$\geq$\,1.2\,mJy AS2UDS sample is shown in Fig.~\ref{fig:pz}b. The 450-$\mu$m population has on average $\sim$\,14 times (at 12-$\sigma$ significance) higher space density up to a redshift of $z$\,$\sim$\,2, but a similar space density at $z$\,$\sim$\,2--3 as the $S_{850}$\,$\geq$\,1.2\,mJy 850-$\mu$m population. We test whether the two distributions are significantly different by using a $\chi^2$ test to compare the space density values at each bin including the errors. The $\chi^2_{\rm red}$\,=\,10 indicates that the two distributions are significantly different. This suggests that 450-$\mu$m-detectable LIRGs are the main obscured population at $z$\,$\sim$\,1--2, while ULIRGs (which make up the samples selected at 850-$\mu$m at the flux limits probed here) dominate at higher redshifts (see \S~\ref{sec:farir}). We further discuss the physical properties of both population at $z$\,$\sim$\,2--3, where the space density is comparable, in \S~\ref{sec:optical}.

%
%
\begin{figure*}
  \includegraphics[width=\textwidth]{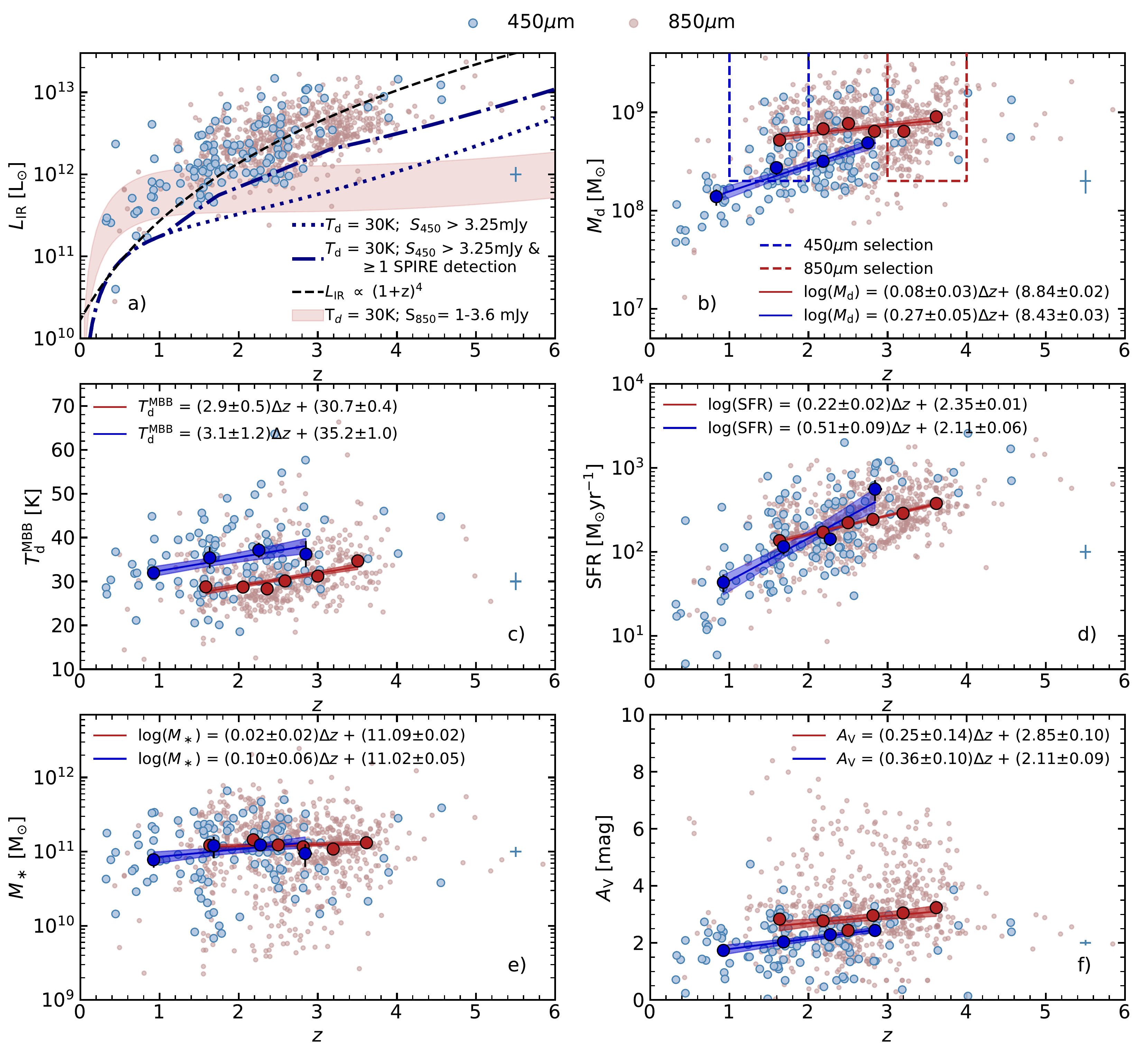}
  \caption{The evolution of the physical properties of the 450-$\mu$m (in blue) population with redshift. In panels b--f, the large circles show the binned median in bins with equal numbers of sources,  the solid line shows the best-fit line around the median redshift (where $\Delta z$\,$=$\,$z-z_{\rm med}$, $z_{\rm med}$\,$=$\, 1.85 and $z_{\rm med}$\,$=$\, 2.61 for the 450-$\mu$m and 850-$\mu$m samples, respectively) to the binned data and the shaded regions show the associated uncertainty. The 850-$\mu$m sample and median values are similarly shown in red.
  {\bf a)} Far-infrared luminosity ($L_{\rm IR}$) evolution. The black dashed line shows the luminosity evolution according to $L_{\rm IR}$\,$\propto (1+z)^4$. The blue dotted line denotes the selection function of an $S_{450}$\,$\gtrsim$\,3.25\,mJy SMG from a modified black body dust SED with $T_{\rm d}\,$\,$=$\,30K (median characteristic dust temperature of the sample). The blue dot-dashed line shows the selection for $S_{450}$\,$\gtrsim$\,3.25\,mJy, including the requirement that the dust SED includes at least one detection above the flux limits of the available SPIRE observations at 250 or 350\,$\mu$m. The shaded region shows the expected limit for the 850-$\mu$m sample from a modified black body dust SED with $S_{850}$\,$=$\,1--3.6\,mJy and $T_{\rm d}$\,$=$\,30K. We see that the populations and their variations with redshift can be roughly described by these selection bounds.
  {\bf b)} Dust mass evolution. No evolution with redshift is seen in the AS2UDS sample due to negative $K$-correction since the source selection is in the Rayleigh-Jeans tail of the SED at the relevant redshifts, which is dominated by the cold dust. There is evolution in the 450-$\mu$m sample with redshift, with sources at $z\lesssim$\,1.5 having lower dust mass due to the $K$-correction at 450\,$\mu$m. To select a uniform  population for assessing evolution we construct a rest-frame matched subset with $M_d$\,$\geq$\,2\,$\times$\,10$^{8}$\,M$_\odot$ over $z$\,$=$\,1--2 for 450\,$\mu$m over $z$\,$=$\,3--4 for the 850-$\mu$m sample. This comparison is discussed in \S~\ref{discussion}.
  {\bf c)} Dust temperature from the modified black body fits for sources with at least one detection in the SPIRE bands. A positive trend with redshift is seen in both 450-$\mu$m and 850-$\mu$m samples. The 450-$\mu$m sample has a systematically higher characteristic dust temperature than the 850-$\mu$m sample.
  {\bf d)} Star-formation rate evolution. A $\sim$\,5-$\sigma$ trend is observed for the 450-$\mu$m sample, similar to the trend seen in the 850-$\mu$m sample. This is mainly due to the selection limit in far-infrared luminosity with redshift (see panel a).
  {\bf e)} Stellar mass evolution. Both samples have comparable stellar masses over all redshift ranges and no significant evolution with redshift is seen in either. 
  {\bf f)} $V$-band dust attenuation evolution. Only weak evolution is seen in both 450-$\mu$m and 850-$\mu$m samples, with the 850-$\mu$m sample having a systematically higher dust attenuation at all redshifts.
  }
  \vspace{-20pt}
  \label{fig:pars_z}
\end{figure*}

\subsection{Physical properties of 450-\texorpdfstring{$\mu$}m sources} \label{properties}

In this section, we analyse the derived physical properties from {\sc magphys}+photo-$z$ of the 121 SMGs selected at 450\,$\mu$m and their variations with redshift, and compare and contrast these with the equivalent properties of the 850-$\mu$m population.

\subsubsection{Far-infrared properties} \label{sec:farir}

As the majority of the emission from these dusty systems is coming from the far-infrared, we begin by investigating the dust properties of the SMGs by deriving their far-infrared luminosities. The median far-infrared luminosity of the 450-$\mu$m sample is $L_{\rm IR}$\,$=$\,(1.5\,$\pm$\,0.2)\,$\times$\,10$^{12}$\,L$_\odot$, with a 16--84th percentile range of $L_{\rm IR}$\,$=$\,(0.7--4.7)\,$\times$\,10$^{12}$\,L$_\odot$. 
In comparison, the 850-$\mu$m population has a median far-infrared luminosity of (2.88\,$\pm$\,0.09)\,$\times$\,10$^{12}$\,L$_\odot$. The 850-$\mu$m SMGs have significantly higher far-infrared luminosities than the 450-$\mu$m SMGs primarily due to their brigher effective flux limit. 

In Fig.~\ref{fig:pars_z}a we plot the luminosity evolution with redshift for both 450- and 850-$\mu$m samples. To consider the influence of sample selection at 450-$\mu$m, we also overlay the predicted far-infrared luminosity of a source with a dust SED modelled by a modified black body with a temperature of $T_{\rm d}$\,$=$\,30\,K (the median for this sample) and a 450-$\mu$m flux density of $S_{450}$\,$=$\,3.25\,mJy, which is the 5-$\sigma$ limit of our 450-$\mu$m sample. We overlay a selection function for the same $T_{\rm d}$\,$=$\,30\,K model with the additional constraint that the SED must be detected in at least one SPIRE band at 250 or 350\,$\mu$m. We see that this selection results in an increasing far-infrared luminosity limit at $z$\,$\gtrsim$\,1 for the 450-$\mu$m sample. The trends of far-infrared luminosity with redshift in our survey can be explained by the flux limit of the sample selection. We also indicate the expected completeness limit for the 850-$\mu$m sample in Fig.~\ref{fig:pars_z}a. The limit of the parent SCUBA-2 850-$\mu$m survey is 3.6\,mJy; however deeper ALMA follow-up observations detect sources down to $S_{870}$\,$\sim$\,1\,mJy, hence the true sample limit is somewhere in-between these values. Due to the negative $K$-correction, we see little variation in the far-infrared luminosity limit with redshift for the 850-$\mu$m selection \citep[see][for presentation of $K$-correction effects in different bands]{2014casey}.

The selection trends in Fig.~\ref{fig:pars_z}a, together with the evolution of the far-infrared luminosity function explains the lower redshift distribution of the 450-$\mu$m sources in comparison to 850-$\mu$m selected sources \citep[e.g.][]{2015bethermin}. As shown in Fig.~\ref{fig:pars_z}a, due to the $K$-correction, the luminosity limit for a given 450-$\mu$m flux density increases quickly with redshift, meaning that the far-infrared luminosity limit is $\sim$\,3 times higher at $z$\,$\sim$\,3 than at $z$\,$\sim$\,1. Combined with the steep decrease in the number of sources at the high luminosity end of the luminosity function at low redshifts, this means that the sources at 450-$\mu$m are detected at lower redshifts than the 850-$\mu$m population. 
Due to stronger negative $K$-correction at 850-$\mu$m, the luminosity limit for 850-$\mu$m sources with S$_{850}$\,$\geq$\,3.6\,mJy is nearly constant across $z$\,$=$\,1--6.
Similarly, the luminosity limit for 850-$\mu$m sources with $S_{850}$\,$\geq$\,3.6\,mJy, though nearly constant across $z$\,$=$\,1--6, is higher than the 450-$\mu$m limit at $z$\,$\leq$\,1.5, explaining the lack of sources detected at lower redshift at 850\,$\mu$m.
The different density evolution seen in Fig.~\ref{fig:pz}b, together with the far-infrared luminosity trends in Fig.~\ref{fig:pars_z}a, indicate that luminous infrared galaxies (LIRGs; $L_{\rm IR}$\,$\simeq$\,10$^{11-12}$\,L$_\odot$) are the main obscured population at $z$\,$\sim$\,1--2, while ultra-luminous infrared galaxies (ULIRGs; $L_{\rm IR}$\,$\simeq$\,10$^{12-13}$\,L$_\odot$) are more dominant at higher redshifts \citep{2013magnelli,2014casey}.

As both surveys are sampled past the peak of the far-infrared SED in the Rayleigh-Jeans tail at the redshifts of interest, we examine the dust masses of the 450-$\mu$m SMGs in comparison to the 850-$\mu$m sample to compare the selection effects on the dust mass distributions.
The median dust mass for the 450-$\mu$m sample is $M_{\rm d}$\,$=$\,(3.6\,$\pm$\,0.2)\,$\times$\,10$^{8}$\,M$_\odot$ with a 16--84th percentile range of $M_{\rm d}$\,$=$\,(2.1--8.2)\,$\times$\,10$^{8}$\,M$_\odot$. For comparison, the 850-$\mu$m sample from \citetalias{2020dudz} has a significantly higher median dust mass (due to its brighter effective flux limit) of $M_{\rm d}$\,$=$\,(6.8\,$\pm$\,0.3)\,$\times$\,10$^{8}$\,M$_\odot$. 
The dust mass evolution with redshift is shown in Fig.~\ref{fig:pars_z}b. A positive trend of dust mass with redshift is observed for the 450-$\mu$m sample, but there is no significant trend for the 850-$\mu$m sources. In agreement with the photometric properties of the samples (see \S~\ref{sec:phot props} and Fig~\ref{fig:obs}), our dust mass results in Fig.\,\ref{fig:pars_z}b, suggest that 450-$\mu$m selection is sensitive to lower dust mass sources at lower redshifts. From the dust mass and far-infrared luminosity results in Fig.~\ref{fig:pars_z}a/b, it is clear that selection at 850-$\mu$m results in a selection that primarily traces cold dust mass; however this is less true for the 450-$\mu$m sample where selection is closer to the peak of the dust SED and thus is more affected by the far-infrared luminosity at $z$\,$\gtrsim$\,2 (see Ikarashi et al., in prep.). 

We now investigate the characteristic dust temperatures derived, for simplicity, from optically-thin modified black body fits, which correspond to the peak of the far-infrared emission.
For the 450-$\mu$m sample, this method provides a median characteristic temperature of $T^{\rm MBB}_{\rm d}$=\,33\,$\pm$\,1\,K with positive evolution of the characteristic dust temperature with redshift seen in Fig.~\ref{fig:pars_z}c. However, we stress that this trend is driven by the increase of luminosity with redshift as a result of selection effects (see Fig.~\ref{fig:pars_z}a), which indeed is found in \citet{2020lim}. We note that \citet{2018zavala} found a mean dust temperature of $T^{\rm MBB}_{\rm d}$\,$=$\,47\,$\pm$\,15\,K. The discrepancy is mostly due to the fitting method as \citet{2018zavala} adopted $\beta$\,$=$\,1.6 and assumed the emission becomes optically thin at $\lambda$\,$\geq$\,100\,$\mu$m, which results in $\sim$\,20 per cent higher characteristic dust temperature values.
Finally, for comparison to the 450-$\mu$m SMGs, we overlay a subset of 475 (out of 707) of the 850-$\mu$m SMGs that have at least one SPIRE detection (to ensure more reliable temperature measurements). This 850-$\mu$m subset has a similar trend, with a comparable gradient and a median characteristic dust temperature of $T^{\rm MBB}_{\rm d}$\,$=$\,30.4\,$\pm$\,0.3\,K. At a fixed redshift, the 450-$\mu$m population appears to be $\Delta T^{\rm MBB}_{\rm d}$\,$=$\,6.0$\pm$1.5\,K hotter than the 850-$\mu$m sample (Fig.~\ref{fig:pars_z}c), or their dust emission becomes optically thick at shorter restframe wavelengths.

Finally, we investigate the star-formation rate, as it is best constrained in the far-infrared regime since the UV/optical wavelengths in SMGs are heavily obscured. The current star-formation rate returned by {\sc magphys}+photo-$z$ is defined as the average star-formation history over the last 10\,Myr. 
For the 450-$\mu$m sample we derive a median star-formation rate of SFR\,$=$\,127\,$\pm$\,20\,M$_\odot$\,yr$^{-1}$ with a 16--84th percentile range of SFR\,$=$\,40--500\,M$_\odot$\,yr$^{-1}$. \citet{2018zavala} suggest a similar star-formation rate (derived from $L_{\rm IR}$) of SFR\,$=$\,150\,$\pm$\,20\,M$_\odot$\,yr$^{-1}$, in agreement with our study.
Comparison of the 450-$\mu$m sources to the 850-$\mu$m sample shows that the latter has a median star-formation rate that is significantly higher, SFR\,$=$\,290\,$\pm$\,14\,M$_\odot$\,yr$^{-1}$, as a result of the brighter effective flux density limit. However, the two distributions overlap as the 850-$\mu$m sample has a 16--84th percentile range of SFR\,$=$\,120--580\,M$_\odot$\,yr$^{-1}$.  
For the 450-$\mu$m SMGs we observe a significant (5\,$\sigma$) variation of SFR with redshift (see Fig.~\ref{fig:pars_z}d), which is driven by the variation in the far-infrared luminosity limit of the sample with redshift. The same trend is observed in the 850-$\mu$m sample, and in Fig.~\ref{fig:pars_z}d we show that the SFRs of the two populations overlap at $z$\,$\simeq$\,1.5--2.5. This indicates that the apparently lower median star-formation rate of the 450-$\mu$m sample is primarily due to the selection being weighted towards less active sources at lower redshifts. 

The best linear fits to the binned values in Fig.~\ref{fig:pars_z} indicate possible differences in the dust properties of the two populations, thus we use a K-S statistic to determine whether these results are significant. We select all sources at $z$\,=\,1.5--2.7, to maximise the overlap between the two samples and exclude any evolutionary trends with redshift. The results indicate that the two samples have significantly different far-infrared luminosity ($P$\,=\,0.002), dust mass ($P$\,=\,3\,$\times$\,10$^{-13}$), dust temperature ($P$\,=\,2\,$\times$\,10$^{-9}$), and star-formation rate ($P$\,=\,0.008) distributions. 

\subsubsection{Optical/near-infrared properties} \label{sec:optical}

\begin{figure*}
  \includegraphics[width=\textwidth]{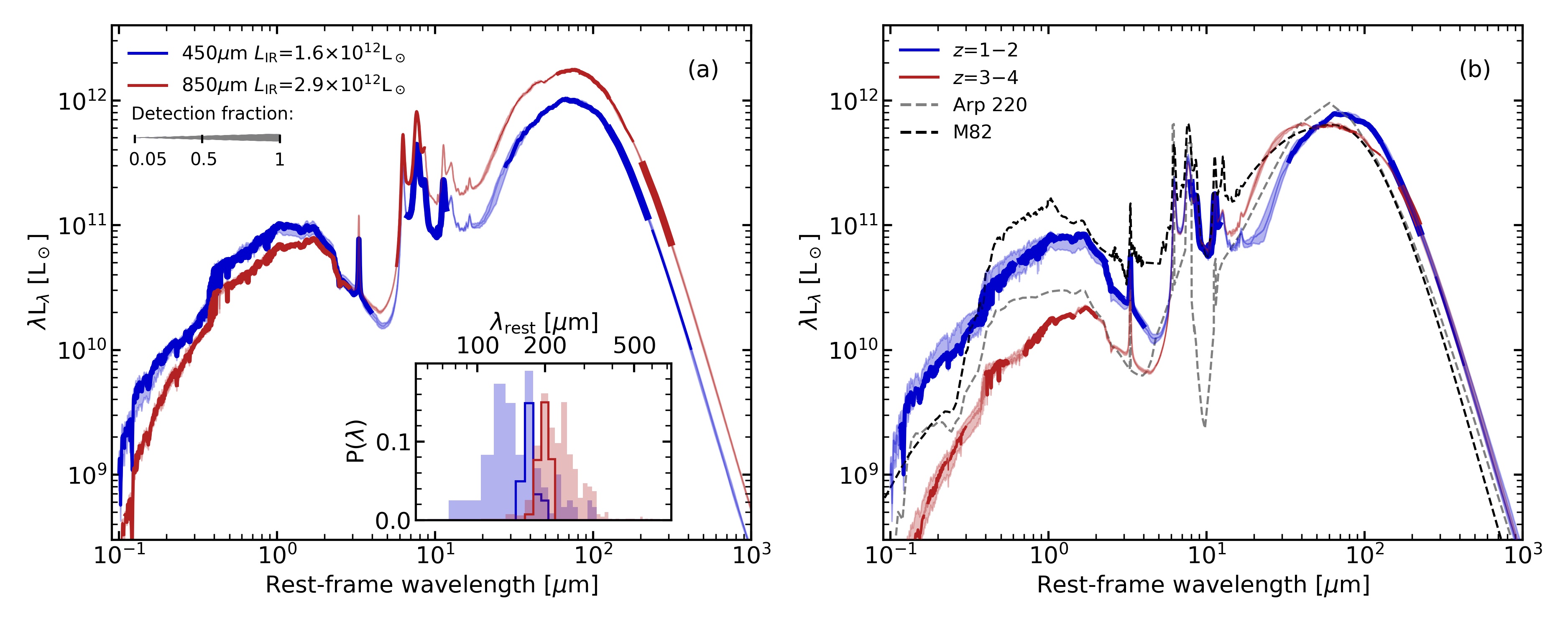}
  \caption{
  {\bf (a)} Median composite SED derived from the best-fit rest-frame SEDs of all 121 STUDIES sources, normalised to the median far-infrared luminosity of the sample. For comparison, we overlay the composite SED of the 850-$\mu$m selected SMGs from the AS2UDS survey, normalised to the median far-infrared luminosity of their sample, $L_{\rm IR}$\,$=$\,2.88\,$\times$\,10$^{12}$\,L$_\odot$. We highlight the reliability of the sections of the SED of each sample with lines of varying thickness corresponding to the detection fraction ($N_{\rm det}$/$N_{\rm sample}$) in each band. Selection at $z$\,$\sim$\,1.5 results in sources that are brighter in the rest-frame optical/near-infrared, likely due to higher stellar masses and/or lower dust attenuation. Bootstrap errors are shown as the shaded regions. The inset panel shows the distributions of rest-frame wavelengths for the 450-$\mu$m and 850-$\mu$m selected samples as shaded histograms in blue and red, respectively. The solid lines indicate the subsets with $M_{\rm d}$\,$\geq$\,2\,$\times$\,10$^8$\,M$_\odot$ at $z$\,$=$\,1--2 and $z$\,$=$\,3--4, corresponding to $\lambda_{\rm rest}$\,$\sim$\,180\,$\mu$m with a $\sim$\,5 per cent deviation.
  {\bf (b)} Median composite SEDs for the  rest-frame-wavelength, $\lambda_{\rm rest}$\,$\sim$\,180\,$\mu$m, matched samples of 31 $z$\,$\sim$\,1.5 sources (selected at observed 450\,$\mu$m) and  220 $z$\,$\sim$\,3.5 sources (selected at observed 850\,$\mu$m), both normalised to the median far-infrared luminosity of the $z$\,$\sim$\,1.5 subset, $L_{\rm IR}$\,$=$\,1.2\,$\times$\,10$^{12}$\,L$_\odot$. Again, we highlight the reliability of the sections of the SED for each sample with lines of variable thickness. For comparison, we also plot the SEDs of the local galaxies M82 and Arp\,220. With similar far-infrared luminosities, the rest-frame near-infrared emission of these two galaxies brackets the $z$\,$\sim$\,1.5 SED and they are both redder in the rest-frame optical, suggesting higher extinction.  On the other hand, the $z$\,$\sim$\,3.5 SED is much fainter in the rest-frame near-infrared, suggesting a lower typical stellar mass or much higher obscuration. In the far-infrared the $z$\,$\sim$\,1.5 subset peaks at a longer wavelength (indicating cooler characteristic dust temperatures or higher opacity) and has a similar width to the dust peak of Arp\,220 (suggesting a similar far-infrared opacity).  In contrast, the $z$\,$\sim$\,3.5 SED peaks at shorter wavelengths owing to the fact that the subsets are matched in rest-frame wavelength and dust mass, and these higher-redshift sources have typically higher far-infrared luminosities and hence are expected to be hotter by $\sim$\,20 per cent.
Bootstrap errors are shown as the shaded regions.}
  \label{fig:composite2}
\end{figure*}

The rest-frame UV/optical/near-infrared features in the SED are dominated by the stellar emission, thus physical properties such as stellar mass and dust attenuation can be inferred. To search for differences in the SED shapes, which also reflect differences in the selection, we stack the rest-frame SEDs of each galaxy (see Fig.~\ref{fig:composite1}b). The SEDs are normalised by their far-infrared luminosity to the median of the sample, $L_{\rm IR}$\,$=$\,1.6\,$\times$\,10$^{12}$\,L$_\odot$, and a composite SED of the whole population is derived by measuring a median value at each wavelength. Fig.~\ref{fig:composite1}b highlights the difficulty of constructing complete samples of strongly star-forming galaxies based on UV/optical observations, as the variation in the SEDs span more than an order of magnitude at restframe wavelengths of $\lambda$\,$\lesssim$2\,$\mu$m (see also \citetalias{2020dudz}). 
The resulting composite 450-$\mu$m SED, together with the equivalent median composite SED of the 850-$\mu$m sources, normalised to $L_{\rm IR}$\,$=$\,2.88\,$\times$\,$10^{12}$\,L$_\odot$ (the median of that sample, \citetalias{2020dudz}), are shown in Fig.~\ref{fig:composite2}a. The error on the median SED is estimated by bootstrap resampling the individual SEDs to form multiple median SEDs and taking the 16th and 84th percentile values at each wavelength.
The shape of the optical SEDs suggest that 450-$\mu$m sources are brighter at $\lambda$\,$\lesssim$\,2\,$\mu$m than the 850-$\mu$m population and indeed, we see in Fig.~\ref{fig:obs}a that the 450-$\mu$m sample has a brighter median observed $K$-band magnitude by 1.3\,$\pm$\,0.2\,mag. This suggests the latter population either has higher stellar masses, lower dust attenuation, and/or younger ages.\footnote{We note, however, that the median sSFR for 850-$\mu$m sources \citepalias{2020dudz} is $\sim$\,0.2\,dex higher than that for 450-$\mu$m sources \citep{2020lim} studied here, suggesting that age is not the main driver of the differences seen in the SEDs.} In the far-infrared, the SEDs, where well constrained, have a similar overall shape, peaking (in $\lambda L_\lambda$) at similar wavelengths, $\lambda_{\rm rest}$\,$\sim$\,80\,$\mu$m. 
As the composite SEDs suggest differences in the physical properties inferred from the optical emission, we next examine whether this is mainly driven by the differences in the stellar masses or the dust attenuation of the two populations.

First, we compare the stellar mass of the two samples. The 450-$\mu$m SMGs have a median stellar mass of $M_\ast$\,$=$\,(1.07\,$\pm$\,0.12)\,$\times$\,10$^{11}$\,M$_\odot$ with a 16--84th percentile range of (0.4--2.3)\,$\times$\,10$^{11}$\,M$_\odot$. In comparison, \citet{2018zavala} find a mean stellar mass of $M_\ast$\,$=$\,(0.99\,$\pm$\,0.06)\,$\times$\,10$^{11}$\,M$_\odot$, which is within $\sim$\,1\,$\sigma$. We see no evolution of stellar mass with redshift, in agreement with the results for the 850-$\mu$m population in the AS2UDS study \citepalias{2020dudz}, as shown in Fig.~\ref{fig:pars_z}e. The median stellar mass of the 850-$\mu$m sample is  $M_\ast$\,$=$\,(1.26\,$\pm$\,0.05)\,$\times$\,10$^{11}$\,M$_\odot$, similar to the 450-$\mu$m population (but typically seen at an earlier epoch).

As both samples have comparable stellar masses but the 450-$\mu$m sample is $\sim$\,0.9\,mag brighter at the rest-frame $V$-band, next we assess whether the differences in the rest-frame optical/near-infrared SEDs is due to the different attenuation of the stellar emission of the two populations. The median $V$-band dust attenuation of the 450-$\mu$m sample is $A_{V}$\,$=$\,2.0\,$\pm$\,0.1\,mag with a 16--84th percentile range of $A_{V}$\,$=$\,1.2--2.9\,mag. The spread at the UV/optical wavelengths seen in Fig.~\ref{fig:composite1}b highlights the variety of the SEDs of the 450-$\mu$m sources, ranging from unobscured (A$_{V}$\,$\sim$\,0)  Lyman-break galaxies, through more typical A$_{V}$\,$\sim$\,1 star-forming galaxies similar to those selected using the $BzK$ criteria \citep{2004daddi}, to higher-A$_{V}$ systems such as  Extremely Red Objects \citep{2004smail} and the near-infrared faint populations \citep{2014simpson,2018franco,2020umehata,2020smail}.

The median dust attenuation is significantly lower than the AS2UDS value of $A_{V}$\,$=$\,2.89\,$\pm$\,0.04\,mag, as also suggested from the comparison of the rest-frame UV slopes in Fig.~\ref{fig:composite2}a.
The variation of $A_{V}$ with redshift is shown in Fig.~\ref{fig:pars_z}f, with a $\sim$\,2-$\sigma$ positive trend. A similar trend is observed in the AS2UDS sample, but offset to higher extinction. Thus, the difference in the optical/far-infrared SEDs is mainly attributed to the lower median value of dust attenuation of the 450-$\mu$m sample compared to 850-$\mu$m sample. 
Since $A_{V}$ is also correlated with far-infrared luminosity, we pick luminosity-matched subsets with $\log_{10}(L_{\rm FIR}/L_\odot)$\,$=$\,12.1--12.7 at $z$\,$=$\,1.5--2.7 from the 450-$\mu$m and 850-$\mu$m samples. We find median dust attenuation of $A_{ V}$\,$=$\,2.29\,$\pm$\,0.13\,mag and $A_{V}$\,$=$\,2.63\,$\pm$\,0.07\,mag, respectively. Both subsamples do not show any significant trend in dust attenuation with redshift, meaning that the difference in dust attenuation between the two samples is most likely due to the variation in the far-infrared luminosity with redshift.

The 450- and 850-$\mu$m samples appear to be different when considering dust properties (see \S~\ref{sec:farir}), therefore we again select all sources at $z$\,$=$\,1.5--2.7 and use a K-S statistic to test whether the trends of optical properties seen in Fig.~\ref{fig:pars_z} are significant. We find that the stellar mass distributions are not significantly different ($P$\,=\,0.3), while the dust attenuation has a probability of the two samples being drawn from the same parent distribution of $P$\,=\,0.0004, indicating a significant difference. 

In Fig.~\ref{fig:pz}b, we show that both samples have a comparable space density at $z$\,$=$\,2--3. We analyse further whether they are similar populations in terms of their physical properties, by selecting all $z$\,$=$\,2--3 sources that have $S_{450}$\,$\leq$\,15\,mJy (using the interpolated value from the best-fit SEDs for the AS2UDS sources) and $S_{850}$\,$\leq$\,4\,mJy, resulting in 26 and 195 sources selected at 450\,$\mu$m and 850\,$\mu$m, respectively. These flux density cuts minimise the overlap between the two populations by selecting sources that would be harder to find at either 450 or 850\,$\mu$m, respectively. We find three main differences: the 450-$\mu$m  sample (those which are harder to find at 850\,$\mu$m) has a higher stellar mass, lower dust attenuation and higher dust temperature compared to the 850-$\mu$m sample. Thus, the same trends as seen in Fig.~\ref{fig:pars_z}, except for stellar mass, remain for the non-overlapping populations. 
Since other physical properties are comparable, we conclude that the $z$\,$=$\,2--3 450-$\mu$m sample has a lower dust-to-stellar mass ratio, suggesting more evolved systems with lower gas fractions.

We also calculate the median infrared excess, $L_{\rm FIR}/L_{\rm UV}$, for both 450 and 850-$\mu$m samples. The UV luminosity is estimated from the rest-frame composite SEDs (see Fig.~\ref{fig:composite2}a) at 1600\AA. We find a median excess of 240$^{+100}_{-25}$ and 1160$^{+180}_{-140}$, respectively, where the errors are calculated from the 16--84th percentile range of the rest-frame SEDs. These ratios are much higher than those for similar stellar mass UV-selected galaxies \citep[$\sim$\,4--90;][]{2013heinis,2016bouwens,2019alvarez}.

The results and the properties discussed in \S~\ref{sec:phot props} and shown in Fig~\ref{fig:obs} show that the 850-$\mu$m population is fainter in the optical regime, with $\sim$\,17 per cent of sources being undetected in $K_{\rm s}$-band, compared to only $<$\,4 per cent for the 450-$\mu$m sources. The $K_{\rm s}$-undetected 850-$\mu$m SMGs reside at higher redshift ($z$\,$=$\,3.19\,$\pm$\,0.08) and have higher dust attenuation dust attenuation ($A_{V}$\,$=$\,5.3\,$\pm$\,0.2\,mag) than the full 850-$\mu$m sample, as highlighted in \citetalias{2020dudz}. We also note that this $K_{\rm s}$-undetected subset has higher stellar mass, with a median of $M_\ast$\,$=$\,(1.48\,$\pm$\,0.07)\,$\times$\,10$^{11}$\,M$_\odot$ and are more luminous, with a median far-infrared luminosity of $L_{\rm IR}$\,$=$\,(3.4\,$\pm$\,0.2)\,$\times$\,10$^{12}$\,L$_\odot$. Other physical properties are comparable to those of the full 850-$\mu$m sample. This suggests that the $K_{\rm s}$-undetected 850-$\mu$m sources have slightly higher star-formation efficiency and lower gas fraction, although within the uncertainty to the $K_{\rm s}$-detected population (see also \citealt{2020smail}). This relative paucity of these extremely dust-obscured sources at 450\,$\mu$m likely reflects the lower sensitivity to the highest redshift and high dust mass sources in the much smaller 450-$\mu$m survey volume (which also makes it harder to detect these rarer extremely dust-obscured sources).

Finally, we look at how the optical properties of the 450-$\mu$m sample compare to that selected in the optical/near-infrared, which detects less active star-forming galaxies (the so-called ``star-forming main-sequence"). For this, we use the $K_s$-band selected UDS field galaxies, which have been analysed in a consistent manner to our 450-$\mu$m sample \citepalias{2020dudz}. We select a subsample of field galaxies with $K_{\rm s}\leq$\,25.3 that  have no contamination flags, and have star-formation rates higher than the 16th percentile value of the 450-$\mu$m sample (SFR\,$\geq$\,41\,M$_\odot$\,yr$^{-1}$), to exclude less-active systems. We also restrict both samples to $z$\,$=$\,1.5--2.7 to exclude any evolutionary effects. We note that $K_s$-band-selected galaxies have over an order of magnitude higher number density compared to the 450-$\mu$m sample in this redshift range. We find that the $K_s$-band sample has a $\sim$\,9 times lower median stellar mass (at 10-$\sigma$ significance), similar dust attenuation ($A_{V}$\,$\sim$\,2.0), slightly lower star-formation rates (2.5-$\sigma$ difference) and a similar median redshift ($z$\,$\sim$\,2.15) to the 450-$\mu$m sample.
These results are in agreement to the findings from the photometric properties of the two populations in Fig.~\ref{fig:obs}. 
Thus, compared to ``normal" star-forming galaxies, the 450-$\mu$m selection detects more massive galaxies with higher dust masses, although this higher dust mass is not reflected in higher dust extinction for their
restframe $\lesssim$\,2--3\,$\mu$m detected stellar continuum emission, as measured by $A_V$.

%
%

\section{Discussion} \label{discussion}

So far, we have investigated the physical properties of the full SNR\,$\geq$\,5 450-$\mu$m-selected sample and compared these to those selected at 850\,$\mu$m. However, as seen in Fig.~~\ref{fig:pars_z}, selection at different wavelengths (in populations whose space density peaks at different redshifts, Fig.~\ref{fig:pz}b) leads to a range of potential selection effects. To assess the  evolution with redshift in physical properties of far-infrared-selected samples, we next exploit the 450-$\mu$m and 850-$\mu$m surveys to construct two
samples matched in terms of selection at rest-frame wavelength, $\lambda_{\rm rest}$\,$\sim$\,180\,$\mu$m.
We achieve this by selecting 450-$\mu$m SMGs in the redshift range of $z$\,$=$\,1--2 and 850-$\mu$m SMGs at $z$\,$=$\,3--4, shown in Fig.~\ref{fig:composite2}b. We note that the median rest-frame wavelength for the samples at $z$\,$=$\,1--2 and $z$\,$=$\,3--4 differs by $\sim$\,5 per cent, but we confirm that precisely matching the redshift distributions to achieve perfect agreement in
their median wavelengths does not change our results.
As seen in Fig.\,\ref{fig:pars_z}b, the 850-$\mu$m flux limit
corresponds to  a higher dust mass, so we match the samples with a further constraint on both the 450-$\mu$m and 850-$\mu$m subsamples to have dust masses of $M_{\rm d}$\,$\geq$\,2\,$\times$\,10$^8$\,M$_\odot$ (this selection is shown in Fig.\,\ref{fig:pars_z}b). This results in samples comprising 31 sources  at $z$\,$=$\,1--2 from the STUDIES 450-$\mu$m survey and 220 sources at $z$\,$=$\,3--4 from the AS2UDS 850-$\mu$m survey, which we will refer to as the ``$z$\,$\sim$\,1.5" and ``$z$\,$\sim$\,3.5" samples, respectively, or ``$\lambda_{\rm rest}$\,$\sim$\,180-$\mu$m matched sample" when we discuss the two samples as a whole.
With both samples selected at the same rest-frame wavelength, $\lambda_{\rm rest}$\,$\sim$\,180\,$\mu$m, and occupying the same parameter space in dust mass (roughly equating to sub-millimetre flux limit), we examine whether there are any physical differences between identical far-infrared-selected galaxies as the age of the Universe doubled between $z$\,$\sim$\,3.5 and $z$\,$\sim$\,1.5. We then discuss the implications of these results for the evolution of the dust content in galaxies and thus galaxy evolution as a whole.

\subsection{Comparing rest-frame-selected populations}

 First, we look at the overall properties of our $z$\,$\sim$\,1.5 and $z$\,$\sim$\,3.5 samples by investigating their composite SEDs in Fig.~\ref{fig:composite2}b. The composite SEDs are normalised to the median far-infrared luminosity of the rest-frame $\lambda_{\rm rest} $\,$\sim$\,180-$\mu$m $z$\,$\sim$\,1.5 sample, $L_{\rm IR}$\,$=$\,1.2\,$\times$\,10$^{12}$\,L$_\odot$. 
The errors on the composite SEDs are estimated by resampling the individual SEDs to form 500 sets of 121 SEDs and constructing composite SEDs for each of those sets. The uncertainty is then estimated by taking the 16th and 84th percentile values at each wavelength.
We observe that the far-infrared to optical luminosity ratio, $L_{\rm IR}/L_{\rm opt}$, of the $z$\,$\sim$\,1.5 galaxies is much lower than that of the $z$\,$\sim$\,3.5 population, suggesting that $z$\,$\sim$\,1.5 population has lower dust attenuation and/or higher stellar masses as discussed for the 450-$\mu$m sample in \S~\ref{sec:optical}.

For comparison, we also show in Fig.~\ref{fig:composite2}b the SEDs of the local starburst galaxies M82 and Arp\,220 \citep{1998silva}, normalised to the same far-infrared luminosity. Compared to Arp\,220, the far-infrared to optical ratio, $L_{\rm IR}/L_{\rm opt}$, of the $z$\,$\sim$\,3.5 sources is higher, while that of the $z$\,$\sim$\,1.5 sources is lower. In the far-infrared, the $z$\,$\sim$\,1.5 subset peaks at the longest wavelength (possibly indicating lower characteristic dust temperature and/or higher optical depth) and has a similar peak width as Arp\,220. 
In comparison to M82, we observe that in the optical/near-infrared regime M82 is the brightest (at fixed far-infrared luminosity) and has redder UV/optical colours than either of the $z$\,$\sim$\,1.5 and $z$\,$\sim$\,3.5 samples. In the far-infrared, the $z$\,$\sim$\,3.5 population appears to peak at a similar wavelength to M82 and has the broadest far-infrared SED.
The broader SED could be the result of a broader distribution of dust temperatures at $z$\,$\sim$\,3.5 or differences in the dust opacity of the two samples, with the $z$\,$\sim$\,3.5 sample potentially having lower dust optical depth. However, we note that the constraints near the peak of the dust SED, especially towards shorter wavelengths, for the $z$\,$\sim$\,3.5 sources are weak and thus uncertain (see Fig~\ref{fig:composite2}). Hence, the broader far-infrared SED may be simply due to these weakly constrained mid-infrared SEDs, where the detection rate in the PACS filters is low, with only 14 (6 per cent) of the 850-$\mu$m sources detected at 100\,$\mu$m and/or 160\,$\mu$m. We find that the composite SED of the sources with a detection in at least one PACS band produces SEDs that unsurprisingly peak at shorter wavelengths, while those SEDs constrained only by limits peak at longer wavelengths; thus when these two 
groups are combined this produces the broad SED. Overall, we conclude that the $z$\,$\sim$\,1.5 sources have properties lying between those of the local templates of Arp\,220 and M82, while the $z$\,$\sim$\,3.5 sources are more extreme than  Arp\,220 in terms of their low rest-frame optical to far-infrared
luminosity ratios.

%
%
\begin{figure*}
  \includegraphics[width=\textwidth]{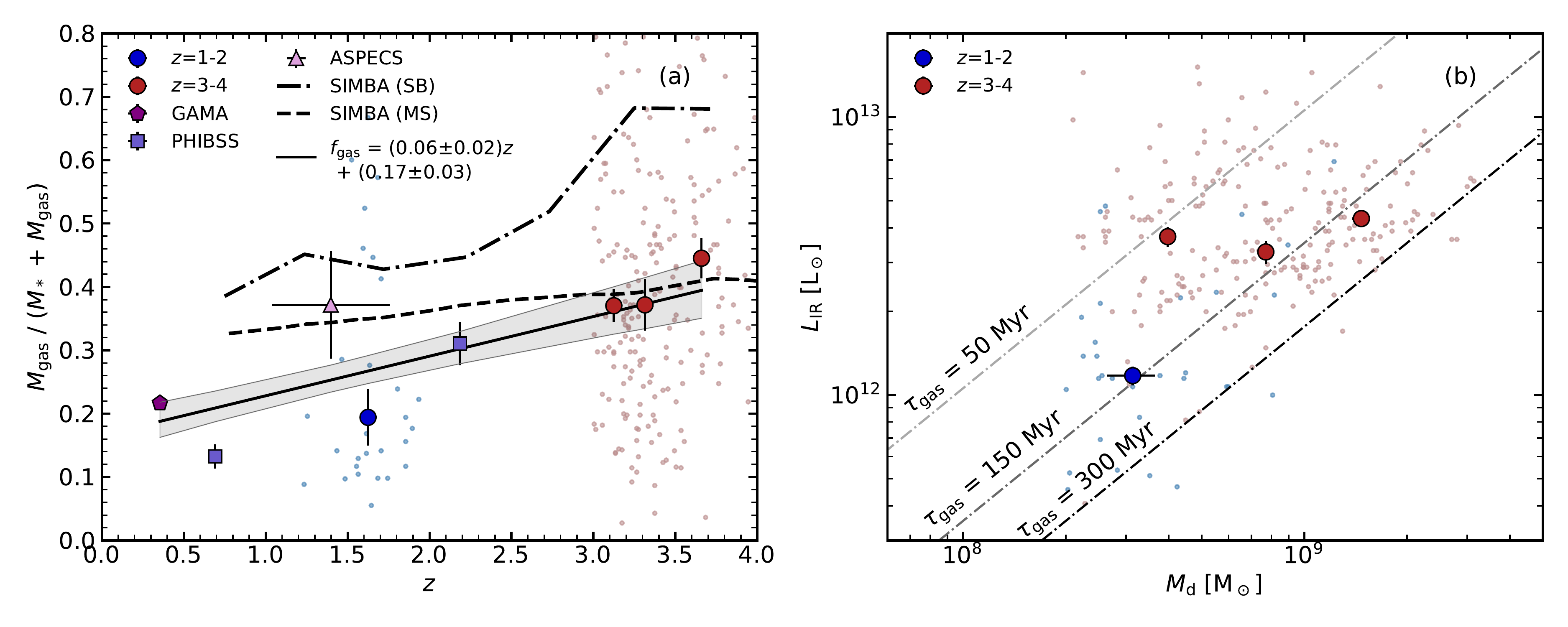}
  \caption{ Results for $z$\,$\sim$\,1.5 and $z$\,$\sim$\,3.5 rest-frame 180-$\mu$m matched samples. 
  {\bf (a)} Gas fraction as a function of redshift. We derive a median gas mass fraction of $f_{\rm gas}$\,$=$\,0.19\,$\pm$\,0.06 with a 68th percentile range of $f_{\rm gas}$\,$=$\,0.10--0.58, assuming a gas-to-dust ratio of 100. We also show results from a sample of $z$\,$<$\,1 ULIRGs from the GAMA survey from \protect\cite{2018driver}, the ASPECS blind CO-survey from \protect\cite{2019aravena} and the
CO-detected typical star-forming galaxies at $z$\,$\sim$\,1--3 from \protect\cite{2018tacconi}. For comparison, we also overlay theoretical predictions for main-sequence (MS) and starburst (SB) galaxies from the {\sc simba} simulations by \protect\citet{2019dave}. The solid line shows the fit to the combined observational data, indicating a modest 3-$\sigma$ increase in gas fraction with redshift, with the uncertainty shown as the shaded region.
{\bf (b)} Far-infrared luminosity as a function of dust mass, the ratio of which is a proxy for star-formation efficiency (or the inverse of gas depletion timescale). The binned median values are shown as large circles, where we split the larger
$z$\,$\sim$\,3.5 sample into three independent bins of dust mass, and the errors are derived by a bootstrap method.  Lines of constant gas depletion are indicated. On average, the star-formation efficiency at $z$\,$\sim$\,1.5 is comparable to that at $z$\,$\sim$\,3.5, but at a fixed dust (and hence gas) mass the higher redshift sources have $\sim$\,3 times higher star-formation efficiency.}
  \label{fig:lm_z}
\end{figure*}

To investigate what drives these differences in the shapes of the SEDs seen in Fig.~\ref{fig:composite2}b, we compare the {\sc magphys}+photo-$z$ derived physical properties between the $z$\,$\sim$\,1.5 and $z$\,$\sim$\,3.5 samples. We find that at $z$\,$\sim$\,1.5 the median stellar mass is $M_\ast$\,$=$\,(1.7\,$\pm$\,0.4)\,$\times$\,10$^{11}$\,M$_\odot$, which is marginally higher than the median stellar mass of the $z$\,$\sim$\,3.5 sample, $M_\ast$\,$=$\,(1.20\,$\pm$\,0.06)\,$\times$\,10$^{11}$\,M$_\odot$. Moreover, $z$\,$\sim$\,1.5 sources have lower dust attenuation, with a median of $A_{ V}$\,$=$\,1.91\,$\pm$\,0.16\,mag, compared to a median of $A_{V}$\,$=$\,3.25\,$\pm$\,0.11\,mag for the $z$\,$\sim$\,3.5 sources. Thus, the brighter optical SED of the $z$\,$\sim$\,1.5 sample arises from the combination of both slightly higher stellar mass and lower dust attenuation. In the far-infrared, we see that the median characteristic dust temperature of $T_{\rm d}$\,$=$\,31\,$\pm$\,3\,K for the $z$\,$\sim$\,1.5 sources is lower, but is consistent within the uncertainties to that of the $z$\,$\sim$\,3.5 SMGs (with at least one SPIRE detection), $T_{\rm d}$\,$=$\,34\,$\pm$\,1\,K. The $z$\,$\sim$\,1.5 population also has a lower median far-infrared luminosity, $L_{\rm IR}$\,$=$\,(1.17\,$\pm$\,0.14)\,$\times$\,10$^{12}$\,L$_\odot$, and dust mass, $M_{\rm d}$\,$=$\,(3.1\,$\pm$\,0.5)\,$\times$\,10$^{8}$\,M$_\odot$, compared to the $z$\,$\sim$\,3.5 population, which has median values of $L_{\rm IR}$\,$=$\,(3.89\,$\pm$\,0.18)\,$\times$\,10$^{12}$\,L$_\odot$ and $M_{\rm d}$\,$=$\,(7.7\,$\pm$\,0.6)\,$\times$\,10$^{8}$\,M$_\odot$, respectively.

\subsubsection{Gas fraction and star-formation efficiency}

Our analysis suggests that at $z$\,$\sim$\,1.5 far-infrared selected galaxies are different to those at $z$\,$\sim$\,3.5, in both the far-infrared and optical regimes, even when selected at the same rest-frame wavelength and the same dust mass limit. To test how these differences link to the physical properties of the populations, we next compare the available fuel the two populations have for star formation and how efficiently this fuel is used  by calculating the gas fraction and the star-formation efficiency for both samples.

We begin by estimating the gas masses from the dust masses
and assuming a gas-to-dust mass ratio, $\delta_{\rm gdr}$. We explore two approaches to determine the appropriate value for $\delta_{\rm gdr}$. Firstly, we just use an empirical estimate of $\delta_{\rm gdr}$\,$=$\,100, since similar values have been derived both for a small sample of high-redshift SMGs with CO(1--0) observations \citep[see][]{2014swinbank} and for Arp\,220 \citep{2011rangwala}. A gas-to-dust ratio of 100 is also considered to be the average value for most local, metal-rich galaxies \citep[e.g][]{2007draine,2014remy} and SMGs are expected to be metal-rich due to their high stellar mass to star-formation rate ratios \citep{2010mannucci}.
In addition, as gas-to-dust mass ratio is expected to vary with stellar mass and redshift, we can also estimate the expected dust-to-gas ratios for the median stellar mass at the median redshift of each sample. We follow \citet{2015genzel} using mass-metallicity relations appropriate for each redshift and find a metallicity dependent gas-to-dust ratio with a fitting formula from \citet{2011leroy}, who fit local star-forming galaxies. We find that the gas-to-dust mass ratios for both $z$\,$\sim$\,1.5 ($\delta_{\rm gdr}$\,$=$\,100$^{+260}_{-100}$) and $z$\,$\sim$\,3.5 ($\delta_{\rm gdr}$\,$=$\,130$^{+300}_{-130}$) samples have significant fitting uncertainties as the mass-metallicity relations are not well constrained, but that both are consistent with the empirical estimate. Therefore we adopt a fixed ratio of $\delta_{\rm gdr}$\,$=$\,100 for both samples. We use this value and the measured dust masses of our $z$\,$\sim$\,1.5 and $z$\,$\sim$\,3.5 samples to estimate their gas masses and gas fractions, $M_{\rm gas}/(M_{\rm gas}+M_\ast)$, and show these in Fig.~\ref{fig:lm_z}a.

We find a median gas mass of $M_{\rm gas}$\,$=$\,(3.1$\pm$0.5)\,$\times$\,10$^{10}$\,M$_\odot$ for the $z$\,$\sim$\,1.5 sample and a median gas fraction of $M_{\rm gas}/(M_\ast+M_{\rm gas})$\,$=$\,0.19\,$\pm$\,0.05 with a 16--84th percentile range of $f_{\rm gas}$\,$=$\,0.10--0.58. 
For the $z$\,$\sim$\,3.5 sample we find a median gas mass of $M_{\rm gas}$\,$=$\,(7.7$\pm$0.6)\,$\times$\,10$^{10}$\,M$_\odot$. As seen in Fig.~\ref{fig:lm_z}a, the $z$\,$\sim$\,3.5 sample has a higher median gas fraction of $f_{\rm gas}$\,$=$\,0.40\,$\pm$\,0.02, with a 16--84th percentile range of $f_{\rm gas}$\,$=$\,0.22--0.65.
We also overlay, in Fig~\ref{fig:lm_z}, the results from a similar analysis of a sample of $z$\,$<$\,1 ULIRGs ($L_{\rm IR}\geq$\,10$^{12}$\,L$_\odot$) in the GAMA survey from \cite{2018driver}, as well as the gas fraction derived directly from CO for the ASPECS blind CO-survey from \protect\cite{2019aravena} and the
CO-detected typical star-forming galaxies at $z$\,$\sim$\,1--3 from \protect\cite{2018tacconi} (where we use $\alpha_{\rm CO}$\,=\,2.5 to convert to gas mass). Overall, we see a 3-$\sigma$ trend of increasing gas fraction with increasing redshift. This suggests that one fundamental difference
between the $\lambda_{\rm rest} $\,$\sim$\,180-$\mu$m selected galaxy populations at $z$\,$\sim$\,1.5, compared
to $z$\,$\sim$\,3.5, is that the former are more evolved, with more gas transformed into stars and thus higher stellar masses and lower gas masses and lower gas fractions.
For comparison, we overlay the gas fraction evolution from the cosmological hydrodynamic simulation {\sc simba} \citep{2019dave,2019li} for main-sequence (MS) and starburst (SB) galaxies. The galaxy is assumed to be starburst if the positive offset from the main-sequence is SFR/SFR$_{\rm MS}$\,$\geq$\,4, where SFR$_{\rm MS}$ is the main-sequence star-formation rate at a given redshift.
The SB model predicts higher gas fraction at all redshifts compared to our observed trend for highly star-forming galaxies. The difference may be due to the fact that we assume a constant gas-to-dust ratio of 100, while the simulated values vary between $\sim$\,100--1000. As the estimates of the gas masses in galaxies depend on the assumed ratio, in \S~\ref{sec:dust_prop} we compare the simulations and our results in terms of dust mass, which is more fundamental measurement requiring fewer assumptions.

In Fig.~\ref{fig:lm_z}b we also plot far-infrared luminosity versus dust mass, where dust mass is a proxy for gas mass and the ratio of these quantities corresponds to the star-formation efficiency (equivalent to the inverse of gas depletion timescale), for the $z$\,$\sim$\,1.5 and $z$\,$\sim$\,3.5 samples.
Using the gas mass for our $z$\,$\sim$\,1.5 sample we estimate the gas-depletion timescale, assuming that half of the gas is available to form stars and the other half is expelled \citep{2002Pettini}, $\tau_{\rm dep}$\,$=$\,$(0.5 \times M_{\rm gas})/{\rm SFR}$,
and overlay lines of constant gas depletion timescale in Fig.~\ref{fig:lm_z}b.
We find a comparable median gas depletion timescale for both samples, $\tau_{\rm dep}$\,$=$\,150\,$\pm$\,40\,Myr at $z$\,$\sim$\,1.5 and $\tau_{\rm dep}$\,$=$\,130\,$\pm$\,7\,Myr at $z$\,$\sim$\,3.5. We can estimate the expected lifetime of the current star-formation event as it is twice the gas depletion timescale if we assume to be observing SMGs halfway through the burst. This approach yields lifetimes of 300\,$\pm$\,80\,Myr and 260\,$\pm$\,14\,Myr for the $z$\,$\sim$\,1.5 and $z$\,$\sim$\,3.5 samples, respectively. The results indicate that the star-formation is slower at $z$\,$\sim$\,1.5, while at $z$\,$\sim$\,3.5 the more gas-rich galaxies  are forming stars more rapidly, and so consuming the larger gas reservoirs in a comparable amount of time.
We can compare these lifetimes to the time taken to form the observed stellar mass, $M_\ast/{\rm SFR}$. This crude age estimates results in a median of 900\,$\pm$\,200\,Myr for the $z$\,$\sim$\,1.5 sample and 400\,$\pm$\,20\,Myr for the $z$\,$\sim$\,3.5 sample. In a simple model where the galaxies are seen on-average halfway through their current star-formation event, the higher formation ages
derived from $M_\ast/{\rm SFR}$, compared to their gas
depletion timescales, suggest that galaxies in both samples
had  pre-existing stellar populations before the onset of the current star-formation event, with those in the $z$\,$\sim$\,1.5 systems being more substantial. 

As the median stellar masses of both samples are comparable, we would expect the lower redshift sample to either have similar metallicities \citep{2013stott} or higher metallicities if the mass-metallicity has dependence on redshift \citep{2015genzel}, which in turn should result in comparable or slightly higher dust attenuation. However, we see the opposite trend, with the $z$\,$\sim$\,1.5 sources having lower dust attenuation than the $z$\,$\sim$\,3.5 population. 

To analyse whether there is any indication of possible differences in the dust continuum structures, specifically the sizes and the dust densities, of the populations at different redshifts, we compare our results to an optically-thick model of the dust emission by \citet{2013scoville}. In this model, the dust cloud is parameterised by a radial power law density distribution with $r^{-1}$ (note, that \citealt{2013scoville} obtained similar results for other reasonable power laws) and a temperature profile combining optically thin emission (with $T_{\rm d}$\,$\propto$\,$r^{-0.42}$) in the central regions ($r$\,$\lesssim$\,1\,pc), optically thick emission (with $T_{\rm d}$\,$\propto$\,$r^{-0.50}$) at intermediate radii and optically thin emission at large radii ($r$\,$\gtrsim$\,2\,kpc). The properties of a far-infrared source are defined by the far-infrared luminosity of the central heating source and the dust mass in the surrounding envelope. \citet{2013scoville} compute emergent spectra for a source of central luminosity of 10$^{12}$\,L$_\odot$ and dust masses ranging 10$^{7-9}$\,M$_\odot$, which are representative of the properties of ULIRGs and SMGs. \citetalias{2020dudz} showed that the 850-$\mu$m SMGs are broadly consistent with this homologous and homogeneous population model of centrally-illuminated dust clouds, with the dust continuum size of SMGs broadly following the expected trend with  far-infrared luminosity-to-gas mass ratio. In Fig.~\ref{fig:lm_z}b we see that, on average, both $z$\,$\sim$\,1.5 and $z$\,$\sim$\,3.5 populations have comparable $L_{\rm IR}/M_{\rm d}$ ratios, hence, if we assume that the sources in our $z$\,$\sim$\,1.5 and $z$\,$\sim$\,3.5 samples can be modelled as having broadly similar structures for the dust continuum regions, they are expected to have comparable effective dust continuum emission radii of $\sim$\,0.8\,kpc. The lower dust density at $z$\,$\sim$\,1.5, due to lower dust mass but comparable dust continuum sizes, may explain the lower dust attenuation compared to $z$\,$\sim$\,3.5 population. 

We note that the median dust mass for our $z$\,$\sim$\,3.5 sample is, on average, two times higher than for the $z$\,$\sim$\,1.5 population. Thus, to check if dust density is still lower at $z$\,$\sim$\,1.5 for galaxies at a given dust mass, we restrict our analysis to have comparable dust masses of $M_{\rm d}$\,$\simeq$\,(2--5)\,$\times$\,10$^{8}$\,M$_\odot$, resulting in 23 sources at $z$\,$\sim$\,1.5 and 61 sources at $z$\,$\sim$\,3.5. We find that the $z$\,$\sim$\,1.5 sources have $\sim$\,3 times lower far-infrared luminosity to gas mass
ratios, $L_{\rm IR}/M_{\rm gas}$. This implies that, for a given dust mass, the $z$\,$\sim$\,1.5 sources potentially have approximately two times larger dust continuum sizes. Thus, $z$\,$\sim$\,1.5 sources have lower dust densities than the $z$\,$\sim$\,3.5 population, due to their larger dust continuum sizes for a given dust mass. Both dust-mass-limited and dust-mass-matched samples suggest that the lower dust density at $z$\,$\sim$\,1.5 is the key parameter leading to lower dust attenuation compared to the $z$\,$\sim$\,3.5 sample.

\subsection{Dust properties of far-infrared-selected galaxies} \label{sec:dust_prop}

Dust, while a small component of the overall baryonic mass of a galaxy, is a useful tracer of the ISM. Therefore, measuring dust mass, especially at different cosmic epochs, is important for understanding the evolution of the ISM in galaxies.
To examine how the dust mass in galaxies has evolved we construct the dust mass function and derive the dust mass density for our $\lambda_{\rm rest}$\,$\sim$\,180-$\mu$m matched samples at $z$\,$\sim$\,1.5 and $z$\,$\sim$\,3.5.

\subsubsection{Dust mass function}

\begin{figure*}
  \includegraphics[width=\textwidth]{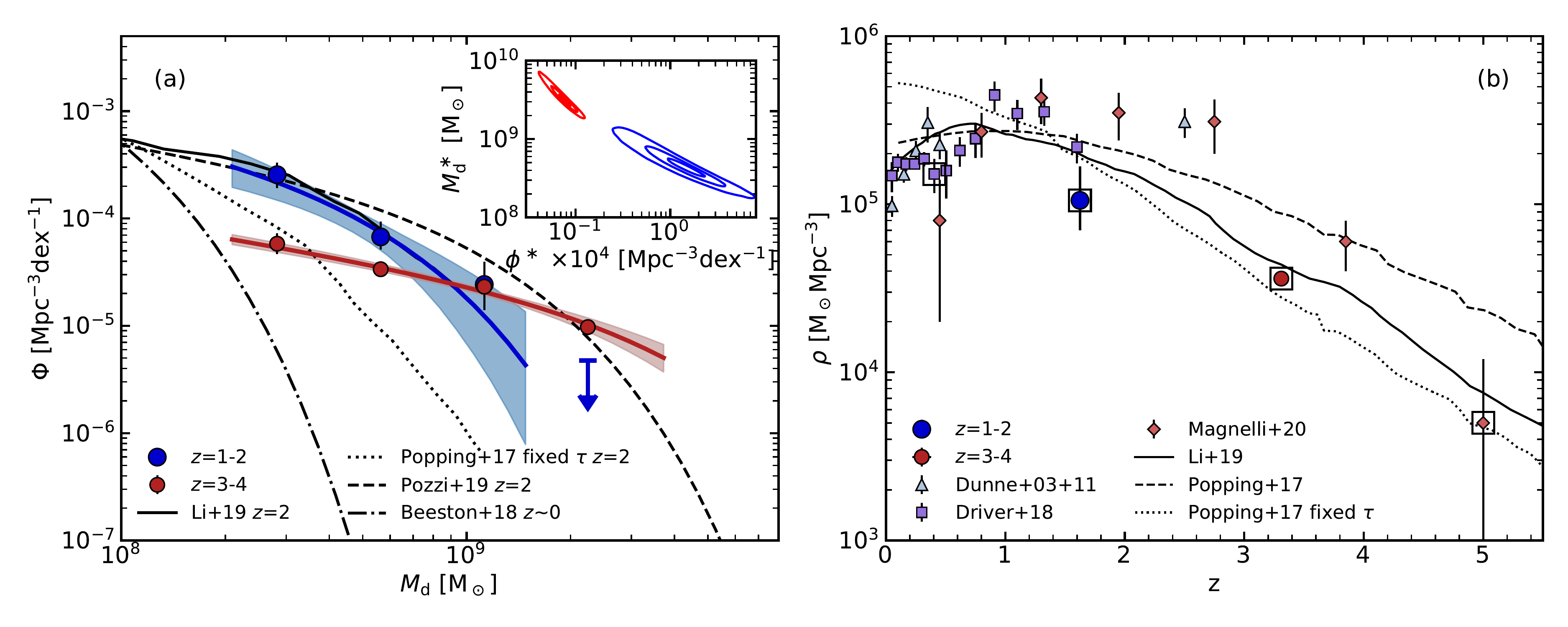}
  \caption{
  {\bf (a)} Dust-mass function for the $\lambda_{\rm rest}$\,$\sim$\,180-$\mu$m matched $z$\,$\sim$\,1.5 and $z$\,$\sim$\,3.5 samples. The arrow indicates a 2-$\sigma$ upper limit for the $z$\,$\sim$\,1.5 sources. For comparison, we overlay the local ($z$\,$<$\,0.1) dust mass function from the GAMA sample of \citet{2018beeston} and $z$\,$=$\,2 results from 160-$\mu$m survey by \citet{2020pozzi}. We also show $z$\,$=$\,2 predictions from a semi-analytical model from \citet{2017popping} assuming a fixed timescale for dust accretion in the ISM of 100\,Myr, and from the cosmological hydrodynamic simulation {\sc simba} by \citet{2019li}. The difference in the shape of our two dust mass functions indicates that the characteristic dust mass of the high-redshift sources is higher than that of the low-redshift sample. Additionally, the number density is higher at lower redshift for all but the highest dust mass sources. The inset panel shows the correlation between the best-fitting Schechter function parameters, characteristic dust mass and space density, for the two samples in their respective colours. Both characteristic dust mass and space density evolve, thus the two $\lambda_{\rm rest}$\,$\sim$\,180-$\mu$m-matched samples do not overlap in this parameter space.
{\bf (b)} Dust mass density as a function of redshift for the rest-frame wavelength $\lambda_{\rm rest}$\,$\sim$\,180-$\mu$m matched $z$\,$\sim$\,1.5 and $z$\,$\sim$\,3.5 samples showing a decline in these matched samples towards higher redshifts. For comparison we overlay results for the total co-moving dust mass densities in galaxies from \citet{2003dunne,2011dunne}, \citet{2018driver} and \citet{2020magnelli}. To highlight the results with a comparable selection, we indicate the rest-frame $\lambda_{\rm rest}$\,$\sim$\,180-$\mu$m selection with black squares.
We also show the predictions from a semi-analytical model from \citet{2017popping}, assuming a fixed timescale for dust accretion in the ISM of 100\,Myr, which we label ``fixed $\tau$", and another model with no dust accretion. Finally, we plot the predicted dust mass densities from the cosmological hydrodynamic simulation of \citet{2019li} and find that it fits both the low redshift samples and our high redshift observations reasonably well.}
  \label{fig:dust_z}
\end{figure*}

We calculate the dust mass function for the $z$\,$\sim$\,1.5 sources using an accessible volume method: $\phi(M_{\rm d})\Delta M_{\rm d}$\,$=$\,$\Sigma (1/V_i)$, where $\phi(M)\Delta M$ is the number density of
sources with dust masses between $M$ and $M$+$\Delta M$ and $V_i$ is
the co-moving volume within which the $i$-th source would be detected
in a given dust mass bin. For each source, the area within which each would have been selected at $\geq$\,5$\sigma$, given its 450-$\mu$m flux density, is calculated in the same manner as in \S~\ref{sec:redshift}. We show the resulting dust mass function in Fig.~\ref{fig:dust_z}a. We indicate a 2-$\sigma$ limit (corresponding to 2.5 sources) in the highest dust mass bin, which has no detected sources. For the $z$\,$\sim$\,3.5 sample (derived from the 850-$\mu$m AS2UDS sources), we correct for incompleteness at faint flux densities by measuring the number of sources to the expected number counts using the slope of the ALMA 1.13-mm counts in the GOODS-S field from \citet{2018hatsukade}. 
We overlay the resulting dust mass function of the $z$\,$\sim$\,3.5 sources in Fig.~\ref{fig:dust_z}a. The uncertainties on the dust functions of both samples were calculated by resampling the dust mass and redshift probability distributions to construct multiple dust mass functions. The resulting error bars are taken as the 16th and 84th percentiles of each bin. 

%
%

To model the mass functions, we fit them using Schechter functions of the form, $\phi $\,$=$\,$ (\phi^\ast/M_{\rm d}^\ast)(M_{\rm d}/M_{\rm d}^\ast)^\alpha e^{-M_{\rm d}/M_{\rm d}^\ast}$, where  $\phi^\ast$ is the characteristic space density, $M_{\rm d}^\ast$ is the characteristic dust mass and $\alpha$ is the power-law slope \citep{1976schechter}. The power-law slope defines the shape of the function at low dust masses and previous studies have yielded values ranging $\alpha $\,$=$\,$-$1.0 to $-$1.7 \citep{2005vlahakis,2011dunne,2013clemens}. As our rest-frame $\lambda_{\rm rest}\simeq$\,180-$\mu$m matched samples are selected to have dust masses of $M_{\rm d}$\,$\geq$\,2\,$\times$\,10$^{8}$\,M$_\odot$, we are unable to constrain $\alpha$ directly and so instead we choose to fix it to $\alpha$\,$=$\,$-$1.5. The Schechter fits to both the $z$\,$\sim$\,1.5 and $z$\,$\sim$\,3.5 samples are shown in Fig.~\ref{fig:dust_z}a.

The best fit for the $z$\,$\sim$\,1.5 sample has $M_{\rm d}^\ast$\,$=$\,3.9$^{+3.3}_{-1.5}$\,$\times$\,10$^{8}$\,M$_\odot$ and $\phi^\ast$\,$=$\,1.6$^{+2.4}_{-1.0}$\,$\times$\,10$^{-4}$\,Mpc$^{-3}$\,dex$^{-1}$, while the best fit for the $z$\,$\sim$\,3.5 sample has $M_{\rm d}^\ast$\,$=$\,3.2$^{+1.6}_{-0.8}$\,$\times$\,10$^{9}$\,M$_\odot$ and $\phi^\ast$\,$=$\,7.6$^{+2.4}_{-2.1}$\,$\times$\,10$^{-6}$\,Mpc$^{-3}$\,dex$^{-1}$. 
In Fig~\ref{fig:dust_z}a, we show the co-variance of the best-fitting Schechter function parameters for both samples. 
The $\chi^2$ contours for the two samples do not overlap, indicating evolution in both characteristic dust mass and space density.
As these subsets are rest-wavelength matched, this change in the shape of their dust mass function suggests a change of normalisation and characteristic dust mass of galaxies with redshift, similar to the findings of \cite{2020pozzi}. However, since we have only limited constraints on the $z$\,$\sim$\,1.5 function due to the small sample size, we caution that the  uncertainties can be substantial.  

For comparison, we overlay the ``local" dust mass function of $z$\,$<$\,0.1 galaxies from the GAMA sample by \citet{2018beeston}. We also show the $z$\,$=$\,2 results from a 160-$\mu$m survey by \cite{2020pozzi}, though we note that at this redshift  their selection wavelength is $\sim$\,50$\mu$m and their survey is thus sensitive to relatively hot dust,
potentially including AGN-heated sources.
We observe that low-redshift sources have higher space density at lower dust masses, but the space density decreases steeply with increasing dust mass. The space density for the low dust mass galaxies is highest at low redshift, while the space density at the high-mass end is higher for the $z$\,$\sim$\,3.5 sources.  
The dust mass functions in Fig.~\ref{fig:dust_z}a suggest that the characteristic dust mass of the $z$\,$\sim$\,3.5 sources is higher than that of the $z$\,$\sim$\,1.5 sample, indicating evolution of a factor 8\,$\pm$\,5 in the characteristic dust mass between $z$\,$\sim$\,1.5 and $z$\,$\sim$\,3.5. 
The normalisations of the best-fit Schechter function suggest that the space density is higher at $z$\,$\sim$\,1.5; however there is an indication that the space density of the highest dust mass sources is higher at $z$\,$\sim$\,3.5. 

We compare our dust mass functions to predictions from a semi-analytical model by \cite{2017popping} and hydro-dynamical simulations by \cite{2019li}, both at $z$\,$=$\,2. We see that \cite{2017popping} model under predicts the observational data at all dust masses, while \cite{2019li} provides a
rough match to our observations at $z$\,=1--2. A description of the models and how they compare to observational results is presented in the subsequent section.

\subsubsection{Dust mass density}

Given the apparently different shapes of the dust
mass function in our two $\lambda_{\rm rest}$\,$\sim$\,180-$\mu$m-matched samples, we opt
to assess the  evolutionary differences in the dust properties of galaxy populations using integrated dust mass density
as the most robust measurement available.
To obtain the dust mass density we use our dust mass measurements together with the accessible volume, which we have calculated in \S~\ref{sec:redshift}.
We derive a dust mass density of $\rho$\,$=$\,(2.6\,$\pm$\,0.5\,)\,$\times$\,10$^{4}$\,M$_\odot$\,Mpc$^{-3}$ at $z$\,$\sim$\,1.5 and $\rho$\,$=$\,(2.41\,$\pm$\,0.13\,)\,$\times$\,10$^{4}$\,M$_\odot$\,Mpc$^{-3}$ at $z$\,$\sim$\,3.5, for a sample with $M_{\rm d}$\,$\geq$\,2.0$\times$\,10$^{8}$\,M$_\odot$. This indicates that galaxies selected at the same rest-frame wavelength  have a similar dust mass density (above a dust mass of $M_{\rm d}\geq$\,2\,$\times$\,10$^{8}$\,M$_\odot$) at $z$\,$\sim$\,1.5 and $z$\,$\sim$\,3.5, assuming that the dust properties used to estimate the dust masses are similar at low and high redshift.

To derive the {\it total} dust mass density, needed to compare to estimates from other studies, we have to extrapolate and integrate the best Schechter fit of each of our
samples from our current limit of $M_{\rm d}$\,$\geq$\,2.0$\times$\,10$^{8}$\,M$_\odot$
down to $M_{\rm d}$\,$=$\,10$^4$\,M$_\odot$, we then add this to the dust mass density of those sources with $M_{\rm d}\geq$\,2\,$\times$\,10$^{8}$\,M$_\odot$, which is calculated above. In this manner, the final total dust mass density we derive for the $z$\,$\sim$\,1.5 sample is $\rho$=\,(1.1$^{+0.6}_{-0.4}$)\,$\times$\,10$^{5}$\,M$_\odot$\,Mpc$^{-3}$, while the $z$\,$\sim$\,3.5 dust mass density is $\rho$=\,(3.6$^{+0.3}_{-0.2}$)\,$\times$\,10$^{4}$\,M$_\odot$\,Mpc$^{-3}$. 
The uncertainties on the dust mass density are a combination in quadrature of the error on the extrapolated values from the dust mass function and the error on the observed dust density (which is obtained by resampling the dust mass and redshift probability distributions).
Thus, the total dust mass density at $z$\,$\sim$\,1.5 is roughly three times higher compared to $z$\,$\sim$\,3.5, at a 2-$\sigma$ level. The larger difference between the extrapolated values is due to the steeper Schechter function for the $z$\,$\sim$\,1.5 subset, as seen in Fig.~\ref{fig:dust_z}a.

Next, we compare our measurements with observations from other studies to assess the dust mass density evolution and investigate the possible physical processes responsible for it. Comparing the dust mass density results is complex due to possible selection effects, as well as small number statistics and uncertainties due to excess variance (sometimes referred to as the cosmic variance). To minimise these, we mainly compare to estimates derived in a similar manner from observations in the far-infrared wavebands.

In Fig.~\ref{fig:dust_z}b, we compare our high redshift measurements to the dust mass density at $z$\,$\simeq$\,0--1 from \citet{2018driver} who use {\sc magphys} to obtain dust masses based on SPIRE photometry for $\sim$\,250,000 galaxies from the GAMA and G10-COSMOS surveys out to $z$\,$\sim$\,1.5. They corrected for volume-limited effects by fitting a spline to the data above their  completeness limits and integrating to obtain total masses. Our $z$\,$\sim$\,1.5 result is within 1.5\,$\sigma$, and thus in agreement with \citet{2018driver} results at the sampled redshift range of $z$\,$\sim$\,1.5. 

We also include results from  a {\it Herschel} SPIRE study at $z$\,$\lesssim$\,0.5 by \citet{2011dunne}, who used {\sc magphys} to estimates dust masses, finding results which agree with the larger subsequent study by \citet{2018driver}.   
As well as a high redshift estimate ($z$\,$\sim$\,2.5) using early SCUBA 850-$\mu$m samples  by \citet{2003dunne}, who fit a modified blackbody with $\beta$\,$=$\,2 (leading to $\sim$\,10 per cent systematic difference in dust mass compared to {\sc magphys}).  We find that their result at $z$\,$\sim$\,2.5 is $\sim$\,2$\sigma$ higher than our estimate at $z$\,$\sim$\,1.5. 

In addition, we overlay the dust mass density results from \citet{2020magnelli}, who analysed 1.2-mm ALMA-selected sources out to $z$\,$\sim$\,5 and calculated the dust mass density by stacking the dust continuum for a $H$-band selected sample, obtaining the total emission for the population. \citet{2020magnelli} fit a modified blackbody with $\beta$\,$=$\,1.8 to obtain the dust mass and calculated the dust mass density, which leads to a systematic difference of $\sim$\,20 per cent compared to fitting it with {\sc magphys}. For this comparison, we overlay their subset with a stellar mass cut of $M_\ast $\,$\geq$\,10$^{9}$\,M$_\odot$, the estimated completeness level of their sample. At $z$\,$\sim$\,3.5 we observe that the dust mass density from \citet{2020magnelli} is marginally higher than our result; however their highest redshift value is consistent with the trend we that see in our subsets. 

We note that the \citet{2003dunne,2011dunne} and \citet{2018driver} samples used SCUBA and {\it Herschel} data,  to constrain their far-infrared SEDs, hence there is additional uncertainty in the identification of counterparts and thus redshift and dust mass estimates, due to source confusion. The ALMA 1.2-mm sample of \citet{2020magnelli} does not suffer from this uncertainty; however, due to the small survey area, the uncertainty in their results arising from excess variance is $\sim$\,45 per cent. We crudely estimate the excess variance in our $z$\,$\sim$\,1.5 sample, by splitting the survey area into independent halves on 0.04\,deg$^2$ scales and derive a mean variance of $\sim$\,44 per cent, which is comparable to the \citet{2020magnelli} findings. For our $z$\,$\sim$\,3.5 sample, which has a larger survey area, this method gives an average excess variance of $\sim$\,12 per cent.

For consistent comparison to our $z$\,$\sim$\,1.5 and  $z$\,$\sim$\,3.5 samples, we highlight the results corresponding to $\lambda_{\rm rest}$\,$\sim$\,180-$\mu$m selection in Fig.~\ref{fig:dust_z}b. We observe that the results follow a smooth trend of decreasing dust mass density with redshift.
The dust mass density evolution allows us to examine possible models of dust formation and growth in galaxies at different epochs. Dust is primarily produced in low/intermediate mass asymptotic giant branch (AGB) stars \citep{1989gehrz,2010sargent} and massive stars at the end of their lives when they explode as supernovae (SNe) \citep{2008rho,2009dunne}. Dust is destroyed by astration, SNe shocks or grain-grain collisions. However, it can reform and grow through accretion in dense and diffuse ISM components. Combining all these processes to predict the lifetime of dust is, therefore, a complicated task, which several groups have attempted to model. Here, we compare our results to the predicted dust mass densities at different epochs from a semi-analytical model by \citet{2017popping} and hydrodynamical simulation {\sc simba} by \citet{2019li}. 
These models differ in several features, most notably \citet{2017popping} consider Type Ia SNe to have the same efficiency in dust formation as Type II SNe, while \citet{2019li} do not consider Type Ia SNe to be significant sources of dust production and omit their contribution from their dust formation model. 

In Fig.~\ref{fig:dust_z}b we overlay the dust mass density for two of the models by \citet{2017popping}. The first assumes that the contribution of dust growth on grains to the dust mass of galaxies is negligible; thus, this model turns off the growth of dust through accretion onto grains and increases the efficiency of dust condensation in stellar ejecta to 100 per cent.
This model predicts more dust than is seen at either $z$\,$\sim$\,1.5 or $z$\,$\sim$\,3.5 in our analysis. Although at low redshift this is only a 1.5-$\sigma$ difference, at $z$\,$\sim$\,3.5 it is a significant $\sim$\,18-$\sigma$ difference, which cannot be accounted for by the excess variance in our sample. 
The second model, which we label ``fixed $\tau$", assumes a fixed timescale for dust accretion in the ISM of $\tau$\,$=$\,100\,Myr, independent of gas density and/or gas-phase metallicity. Although this model matches our
observations within the uncertainties at $z$\,$\sim$\,1.5 and $z$\,$\sim$\,3.5, it does not follow the observational results from other studies  at $z$\,$\lesssim$\,1. At low redshift, it appears that the dominant form of dust production in this model is dust growth by accretion in the ISM since there is no decline at low redshift. The comparison of the model and data suggests that the adopted dust accretion timescale is too short at low redshifts, leading to overestimated dust masses. 

We also compare to predicted dust mass densities from a cosmological hydrodynamic simulation, {\sc simba}, by \citet{2019li}. The predicted dust mass density in their model peaks at $z$\,$\sim$\,1 and declines to the present day, in agreement with the observations. This is due to the decline, on average, of star-formation as a result of the onset of quenching in massive galaxies, which slows down the metal enrichment and limits  grain growth. At higher redshifts, the dust mass density declines steeply, in agreement with our observations at $z$\,$\sim$\,1.5, and slightly overpredicts the dust density at $z$\,$\sim$\,3.5, after taking excess variance into account. This indicates that the grain growth may be weaker or the dust destruction is stronger than assumed in the model, at least at high redshift. In addition, the assumed fixed dust destruction and condensation efficiencies may actually be functions of the local ISM properties. Nevertheless, overall it appears that the assumptions in \citet{2019li}, who combine the dust production by AGB stars and Type II SNe, growth by accretion (with a non-constant accretion timescale) and destruction by thermal sputtering and SNe, may currently provide the best match to observations.

\section{Conclusions}\label{conclusions}

In this paper, we have analysed the physical properties of an effectively completely identified sample of 121 SMGs selected at 450\,$\mu$m from the SCUBA-2 Ultra Deep Imaging EAO Survey \citep[STUDIES;][]{2017wang}. We used {\sc magphys}+photo-$z$ to fit spectral energy distribution models to the available UV-to-radio photometry (a maximum of 24 bands). This allowed us to compare and contrast the population of these 450-$\mu$m selected SMGs to the large sample of ALMA-identified 850-$\mu$m selected sources from AS2UDS (\citealt{2019stach}; \citetalias{2020dudz}), which was analysed in a consistent manner. To investigate how the physical properties of infrared luminous galaxies evolve with redshift, we also select $z$\,$=$\,1--2 450-$\mu$m sources and $z$\,$=$\,3--4 850-$\mu$m sources both with $M_{\rm d}$\,$\geq$\,2\,$\times$\,10$^{8}$\,M$_\odot$, to construct  rest-frame wavelength ($\lambda_{\rm rest}$\,$\sim$\,180\,$\mu$m) matched subsets. We summarise our main findings below. 

\begin{enumerate}[label=(\roman*),align=left]
    
    \item We derive a median redshift of $z$\,$=$\,1.85\,$\pm$\,0.12 for the 121 SMGs selected at 450\,$\mu$m with only $\sim$\,9 per cent lying at $z$\,$\geq$\,3. The distribution can be roughly described by evolution of the far-infrared luminosity function and the 450-$\mu$m flux selection limit.
    The median redshift is significantly lower than the median of the 850-$\mu$m selected sample, $z$\,$=$\,2.61\,$\pm$\,0.08. The fainter 450-$\mu$m sample has, on average, $\sim$\,14 times higher space density than the brighter 850-$\mu$m sample out to $z$\,$\sim$\,2, and a comparable space density at $z$\,$=$\,2--3, before declining. For the $z$\,$=$\,2--3 matched subsets we find that the 450-$\mu$m sources have a lower dust-to-stellar mass ratio, suggesting more evolved systems with lower gas fractions.
    
    \item We find that the 450-$\mu$m sample has a significantly lower dust attenuation of $A_{ V}$\,$=$\,2.1$\pm$\,0.1\,mag, compared to the $A_{V}$\,$=$\,2.89$\pm$\,0.04\,mag of the 850-$\mu$m sample. The SEDs of the 450-$\mu$m population span a wide range in flux at observed optical wavelengths, from unobscured LBGs, through more typical $A_{V}$\,$\sim$\,1 star-forming galaxies, to very obscured and completely optically undetected sources. 
    
    \item The 450-$\mu$m sample has a median dust mass of $M_{\rm d}$\,$=$\,(3.6\,$\pm$\,0.2)\,$\times$\,10$^{8}$\,M$_\odot$, median far-infrared luminosity of $L_{\rm IR}$\,$=$\,(1.5\,$\pm$\,0.2)\,$\times$\,10$^{12}$\,L$_\odot$ and median star-formation rate of SFR\,$=$\,130$\pm$\,20\,M$_\odot$\,yr$^{-1}$, significantly lower than equivalent measures of the 850-$\mu$m sample. These differences are mainly due to the brighter effective flux limit of the 850-$\mu$m sample. 
    
    \item For the rest-frame wavelength-matched, $\lambda_{\rm rest}$\,$\sim$\,180\,$\mu$m, subsets, we find that $z$\,$\sim$\,1.5 subset has a gas fraction of $f_{\rm gas}$\,$=$\,\,0.19\,$\pm$\,0.06. Combined with previous studies, we see a modest 3-$\sigma$ trend of increasing gas fraction with increasing redshift in these populations. Overall,  the galaxies at 
    $z$\,$\sim$\,1.5 and $z$\,$\sim$\,3.5  have comparable star-formation efficiency (although we note that at a fixed dust mass the $z$\,$\sim$\,1.5 sources have $\sim$\,3 times lower star-formation efficiency). Both the lower gas masses and lower star-formation rates at $z$\,$\sim$\,1.5, compared to the $z$\,$\sim$\,3.5 population, lead to comparable remaining lifetimes of the SMG phase of 250--300\,Myr.

    \item By comparing the far-infrared luminosity to gas mass ratios of dust-mass-limited ($M_{\rm d}$\,$\geq$\,2\,$\times$\,10$^{8}$\,M$_\odot$) samples
    at $z$\,$\sim$\,1.5 and $z$\,$\sim$\,3.5, using an optically-thick model by \citet{2013scoville}, we suggest that the $z$\,$\sim$\,1.5 population has lower dust density (assuming similar geometry) due to comparable inferred dust emission radii ($\sim$\,0.8\,kpc) and lower dust mass compared to the $z$\,$\sim$\,3.5 sources. The same is true for dust-mass-matched ($M_{\rm d}$\,$=$\,2--5\,$\times$\,10$^{8}$\,M$_\odot$) samples, as the $z$\,$\sim$\,1.5 sources appear to have lower dust densities due to their  $\sim$\,2 times larger inferred dust continuum sizes compared to the $z$\,$\sim$\,3.5 population. Thus, dust density appears to be a key parameter leading to the lower dust attenuation in SMGs seen at  $z$\,$\sim$\,1.5.
    
    \item We calculate the total dust mass density for the  $\lambda_{\rm rest}$\,$\sim$\,180-$\mu$m-matched samples at $z$\,$\sim$\,1.5 and $z$\,$\sim$\,3.5 by combining the dust mass density estimates extrapolated down to $M_{\rm d} \sim$\,10$^{4}$\,M$_\odot$ from the best-fitting Schechter function for their respective dust mass functions. We find the $z$\,$\sim$\,1.5 sample to have $\sim$\,3 times higher dust density than our $z$\,$\sim$\,3.5 estimate. After combining our results with other far-infrared samples, we find that the model from hydrodynamical simulations by \citet{2019li} combining the dust production by AGB stars and Type {\sc ii} SNe, growth by accretion (with a non-constant accretion timescale) and destruction by thermal sputtering and SNe, is best able to match the observational data. Thus, the dust content in galaxies appears to be governed by a combination of both the variation of gas content and dust destruction timescale.

The comparison of rest-frame wavelength matched samples indicated several potential differences in far-infrared luminous populations as a function of redshift. 
To test the trends uncovered in this study, spectroscopic redshifts based on reliable STUDIES 450-$\mu$m identifications through optical/near-infrared spectroscopy, supplemented by
searches for low-/mid-$J$ CO emission lines in more obscured sources,  will be necessary to confirm the redshift distribution differences between the 450- and 850-$\mu$m populations, as well as to improve the precision of the derived parameters from SED fitting.  Targeting low- and mid-$J$ CO emission lines in these systems would greatly aid in constraining the gas masses and in turn confirm the trends seen in the gas fraction evolution and gas depletion timescales. As this study suggests potential structural differences in the dust continuum structures of far-infrared luminous galaxies at different redshifts, spatially-resolved sub-millimetre interferometry using ALMA is needed to constrain the dust continuum sizes of the 450-$\mu$m population. In addition, future SCUBA-2 450-$\mu$m observations through cluster lenses will uncover populations of even fainter sub-mJy sources, helping to link the populations of dust-obscured and ``normal'' star-forming galaxies at cosmic noon. Such future observations will allow detailed questions to be addressed about the nature of dust and the role of high star-formation rate events in galaxies over a wide range of cosmic epochs.

\end{enumerate}

\section*{Acknowledgements}

The authors thank the anonymous referee for their insightful comments and suggestions that improved the paper. UD acknowledges the support of STFC studentship (ST/R504725/1). The Durham co-authors acknowledge support from STFC (ST/P000541/1 and ST/T000244/1). The James Clerk Maxwell Telescope is operated by the East Asian Observatory on behalf of The National Astronomical Observatory of Japan; Academia Sinica Institute of Astronomy and Astrophysics; the Korea Astronomy and Space Science Institute; Center for Astronomical Mega-Science (as well as the National Key R\&D Program of China with No. 2017YFA0402700). Additional funding support is provided by the Science and Technology Facilities Council of the United Kingdom and participating universities in the United Kingdom and Canada. Additional funds for the construction of SCUBA-2 were provided by the Canada Foundation for Innovation. The submillimetre observations used in this work include the STUDIES program (program code M16AL006). For the near-infrared photometry we use data products from observations made with ESO Telescopes at the La Silla Paranal Observatory under ESO programme ID 179.A-2005 and data products produced by CALET and the Cambridge Astronomy Survey Unit on behalf of the UltraVISTA consortium. HD acknowledges financial support from the Spanish Ministry of Science, Innovation and Universities (MICIU) under the 2014 Ram\'{o}n y Cajal program RYC-2014-15686 and AYA2017-84061-P, the later one co-financed by FEDER (European Regional Development Funds). MPK acknowledges support from the First TEAM grant of the Foundation for Polish Science No. POIR.04.04.00-00-5D21/18-00. LCH acknowledges support from the National Science Foundation of China (11721303, 11991052) and the National Key R\&D Program of China (2016YFA0400702). 

\section{Data Availability}

The data underlying this article are available in the JCMT, ESO, VLA, \textit{Herschel}, \textit{Spitzer}, CFHT and Subaru archives.

\bibliography{studies_bib} 

\bsp	
\label{lastpage}
\end{document}